\documentclass[fleqn,usenatbib]{mnras}

\usepackage{bm}
\usepackage{graphicx}
\usepackage{newtxtext,newtxmath}
\usepackage[T1]{fontenc}
\usepackage{amsmath}

\DeclareRobustCommand{\VAN}[3]{#2}
\let\VANthebibliography\thebibliography
\def\thebibliography{\DeclareRobustCommand{\VAN}[3]{##3}\VANthebibliography}

\title[Circumstellar disks in various binary star configurations]{N-body interactions and collisions in circumstellar disks for planar and inclined binary star configurations}

\author[M. Zimmermann and E. Pilat-Lohinger]{M. Zimmermann$^{1,2}$\thanks{E-mail: maximilian.zimmermann@univie.ac.at}
and E. Pilat-Lohinger$^{2,3,1}$\\
$^{1}$Department of Astrophysics, University of Vienna, Türkenschanzstraße 17, 1180 Vienna, Austria\\
$^{2}$Institut fuer Theoretische Physik - Computational Physics, TU Graz, Petersgasse 16/II, 8010 Graz, Austria\\
$^{3}$Space Research Institute, Austrian Academy of Sciences, Schmiedlstrasse 6, 8042 Graz, Austria\\
}

\date{Accepted 2025 December 9. Received 2025 December 4; in original from 2025 October 27}
\pubyear{2025}

\begin{document}
\label{firstpage}
\pagerange{\pageref{firstpage}--\pageref{lastpage}}
\maketitle

\begin{abstract}
The discovery of exoplanets in binary star systems—now numbering about 850 of the nearly 4,600 known exoplanet systems—raises questions about whether observational bias or stellar companions inhibit planet formation. While most studies on terrestrial planet formation assume planar configurations, wide binaries likely feature random inclinations, potentially disrupting planet-forming disks. This study explores the evolution of embryo-planetesimal disks in S-type motion in misaligned binary systems, focusing on the stage after the gas phase when terrestrial planet formation begins and gravitational interactions dominate. Using our GPU-accelerated N-body code GANBISS, we simulate disks with 2,000 planetesimals and 25 planetary embryos, studying the influence of the planetesimals on the evolution of the embryos and tracking their growth through collisions. After the simulations, we analyze collision outcomes with an analytical model. Moreover, for certain inclined binary configurations, we compare dynamically excited (perturbed by the secondary star) with cold disks in inclined configurations, as the distribution after the gas phase in misaligned binaries remains unclear. Our simulations reveal two key outcomes: (i) embryos migrate slightly inward in misaligned systems, and (ii) The initial large oscillations in embryos’ inclinations and nodes  around the respective values of the secondary star dampen over time. Collision analysis shows distinct differences: planar systems favour accretive collisions, while inclined configurations exhibit more destructive events. These findings underscore the sensitivity of planet formation dynamics to binary star alignment and initial disk conditions.

\end{abstract}


\section{Introduction}\label{intro}

More than 4580\footnote{\url{exoplanet.eu}} planetary systems have been discovered so far. Although most stars are born as part of binary or multiple star systems \citep{DuqMay91,Ragetal10}, only a small fraction of discovered planetary systems have been found within such stellar systems (up to 859\footnote{in binary stars with separations $\leq 2000\ \mathrm{au}$; see \url{https://exoplanet.eu/planets_binary/} and \url{https://exoplanet.eu/planets_binary_circum/}}). This leads to the question if binary systems are unsuitable for planet formation or is this due to an observational bias. On the one hand, radial velocity surveys (RV), which have found about half of the discovered exoplanets in binary systems, often ignore all binary stars with separations $\leq 2''$, as these may fall into the spectrograph slit \citep{EggUdr10,Ngoetal17}. On the other hand transiting objects might be labelled wrongly as single stars. Recently \citet{Suletal24} revisited a sample of Kepler Objects of Interest (KOI) and found 101 new planets, located in binary star systems, which would favour planet formation in binary star systems.\\
In binary stars we distinguish mainly two types of planetary motion \citep[see e.g.~][]{Dvo82, Dvo86}: P-type motion, where a planet orbits both stars, and S-type motion, where a planet orbits only one of the two stars. From the discovered planetary systems, the majority are in S-type motion. In both cases a stellar companion can have a strong effect on the formation and evolution of planets. The strength of the influence strongly depends on the binary parameters (namely, the separation $a_b$, the eccentricity $e_b$, the inclination $i_b$).\\
Planet formation begins on the smallest scale, from sub-micron sized dust to gas giants. Dust particles coagulate into millimetre to centimetre sized agglomerates \citep{Gueetal10,Zsoetal10} due to direct sticking collisions. These centimetre sized particles can be sufficiently concentrated via the gravitational instability \citep{GolWar73,Youetal02} or by streaming instability \citep{YouGoo05,JohYou07,Johetal11} to form planetesimals, which are then thought to collide and form larger planetesimals or planetary embryos. In the end they will result in either terrestrial planets or the cores of giant planets. However, there are still open questions regarding the formation of planets when a second star is influencing this process:\\
(i) The truncation of the circumstellar protoplanetary disk \citep{ArtLub94} affects mainly disks in binary star systems with separations $\leq 150-200\ \mathrm{au}$. Observations show that disks in binary stars with separations less than 40 au, appear less frequent than disks in single star systems \citep{Kraetal12}. Even if a disk is present in a tight binary system, it is less massive. As a consequence, the disk may not contain enough solids to form gas giants or terrestrial planets. Furthermore, the disk is more short-lived, which leaves less time for giant planet formation. \citet{Kraetal12} showed, that the typical lifetime of disks in binary stars with separations $\leq 40\ \mathrm{au}$ is less than $1\ \mathrm{Myr}$, while in a single star system it is $5-10\ \mathrm{Myr}$. Thus, a gas giant has to be formed very quickly.\\
(ii) Dust coagulation; \citet{Zsoetal11} found that due to the perturbation of the secondary star, the eccentricity of the gas disk is pumped up, which increases the relative velocity between the dust and the gas. As a consequence the maximum particle size and the dust coagulation is reduced. This also indicates a smaller average mass of the particles compared to a single star system.\\
(iii) The perturbation of the secondary could be a real problem to the growth of planetesimals through mutual collisions. These are highly sensitive to their collision velocities \citep{BenAsp99,SteLei09,Hep78,Whietal98}.\\
However, \citet{SilRaf21} carried out coagulation-fragmentation model to show that planetesimals can grow in tight S-type orbits in binary systems such as $\alpha$ Centauri or $\gamma$ Cephei starting from $1-10\ \mathrm{km}$. The streaming instability model may also form $100-1000\ \mathrm{km}$ sized planetesimals \citep{Johetal07,Johetal11}. \\
Many investigations have shown that terrestrial planet formation is not suppressed by a secondary star \citep[see e.g.~][]{Baretal02,QuiLis06,Quietal07,Hagray07,Gueetal08,TheHag15,Piletal18}. However, these studies focused mostly on planar binary-disk configurations. But for binary separations $>30 - 40\ \mathrm{au}$ the mutual inclinations of the two stars are randomly distributed \citep{Hal94}.\\
Investigations on planet formation in misaligned binary stars are less common. \citet{Maretal09} and \citet{Xieetal11} have examined the intermediate phase, i.e., from planetesimals to planetary embryos. The latter authors find an inward jumping of planetesimals and piling up in the inner region of the disk. \citet{Quietal02} and \citet{Zhaetal18} investigated the final phase of terrestrial planet formation under the influence of a misaligned secondary star. \citet{Zhaetal18} followed the work of \citet{Xieetal11} and found in the region near the primary star an alignment of the planetary embryos on the inclination of the secondary star. However, both \citet{Quietal02} and \citet{Zhaetal18} started with a planar (dynamically cold) disk.\\
In this study, we assumed initially dynamically excited disk. Furthermore, the total disk mass is not fully contained in planetary embryos due to a considerable fraction of destructive collisions during the gas phase. Thus, a significant fraction of the disk mass consists of planetesimals \citep{Gyeetal14,MarSch00,Paretal08}. Therefore, in the beginning of the final phase of terrestrial planet formation, the disk may consist of a large amount of planetesimals and some planetary embryos. This means that the gravitational interactions between all disk objects have to be taken into account, which significantly increases the computational effort with $O\left(N^2\right)$. CPU N-body codes are generally not suitable for such simulations. Highly parallelized GPU N-body codes have to be taken into account \citep[see][for single star and binary star systems, respectively]{GriSta14,ZimPil23}.\\
In addition to the dynamical evolution of the disk, collisions between disk objects during N-body computations are also an major issue. Typically, the so-called perfect merging approach is used, which might be feasible for low velocity collisions. However, in reality, collisions are not perfect inelastic. More realistic collisions outcomes can be achieved with smooth-particle hydrodynamic (SPH) simulations \citep[see e.g.~][]{BenAsp94,Schetal16,Maietal13,Genetal12}. However, these are complex computations resulting in an increased computation time. On the other hand, \citet{LeiSte12} developed an analytic model for predicting different collision outcomes based on the collision parameters.\\
In this study we performed N-body simulations of planetary embryo-planetesimal disks in S-type motion in various binary star configurations, using our GPU parallelized N-body code GANBISS. As a first approximation we applied the analytic collision model, developed by \citet{LeiSte12}, as a post-processing on the collisions occurred in our N-body simulations, for which we assumed perfect-merging.\\
This paper is structured as follows: In section 2 we describe our numerical models, the initial condition sets and the methods for analysing our results. The results of the dynamical evolution of the disks and those of the collisions are presented in sections 3 and 4, respectively. Finally, section 5 discusses the outcome and provides a summary.

\section{Computations}\label{numerics}
\subsection{Numerical model}

{\it N-body simulations:}\\
Our investigation concerns the stage after the gas phase, so only gravitational interactions are considered. We study the motion of planetesimals and planetary embryos in circumstellar disks in binary stars using our self-developed GPU N-body code GANBISS \citep{ZimPil23}.\\
To calculate the interactions between all bodies (i.e.~the two stars, planetary embryos, and all planetesimals), the equations of motion have to be solved:
\begin{align}
  \dot{\bm{r}}_\nu &= \bm{v}_\nu\\
  \dot{\bm{v}}_\nu &= k^2 \sum^N_{\mu=1,\nu\neq\mu}\frac{m_\mu\left(\bm{r}_\mu - \bm{r}_\nu\right)}{\left\|\bm{r}_\mu-\bm{r}_\nu\right\|^3}
\end{align}
GANBISS uses a GPU parallelized approach which allows the simulation of some thousand interacting objects, and it uses the Bulirsch-Stoer method \citep{StoBul80} with an adaptive step-size. For the sake of computation time, we have chosen a lower error tolerance of $\epsilon=10^{-9}$ for the calculations. The simulations have been carried out on NVIDIA A100 and NVIDIA A40 GPUs. The planetary embryos grow to terrestrial planets via collisions with other disk objects where perfect merging was assumed for these two body collisions.\\

\subsection{Initial conditions}\label{initCond}

{\it Binary star configurations:}\\
We consider equal mass binary star configurations where each star has $1\ \mathrm{M_{\sun}}$. In our study, we varied the following orbital parameters of the binary stars:
(i) the semi-major axis ($a_b$) between $30$ and $100\ \mathrm{au}$,
(ii) the eccentricity ($e_b$) between 0 and $0.4$, and
(iii) the inclination ($i_b$) between 0 and $20{\degr}$ compared planar and inclined stellar configurations. For simplicity, the inclination\footnote{The reference plane for $i_b$ is usually the plane of the embryo-planetesimal disk around the primary star -- unless otherwise specified.} of the secondary star ($i_b$) is varied with respect to the plane of the primary's planetesimal disk. All configurations were chosen such that all disk objects were initially within the stable area defined for S-type motion in planar configurations \citep[see][]{PilDvo02,HolWie99}. The simulation time for each system was $10\ \mathrm{Myr}$. For calculations after this period, one can switch to a CPU code, as the number of planetesimals and thus their influence on the evolution of the embryos is no longer significant.
Due to the variations of the orbital parameters ($a_b,e_b,i_b$) of the binary stars, the different systems are labelled as aXX-eYY-iZZ, where XX indicates the stellar separation, YY the eccentricity, and ZZ the inclination of the binary system (see first column of Table~\ref{tab:binConfTP}).

{\it Circumstellar disk:}\\
All planetesimals and planetary embryos are placed in a ring between $1$ and $4\ \mathrm{au}$ around the primary star. The distribution follows the power law
\begin{equation}
\Sigma\left(r\right) = \Sigma_0\left(\frac{r}{1\ \mathrm{au}}\right)^{-\alpha}
\label{eq:aDist}
\end{equation}
with $\Sigma_0 \approx 1.125 \cdot 10^{-6}\ M_{\sun}\ \mathrm{au}^{-2}$ and $\alpha=1.5$, i.e.~the usual value for the Minimum Mass Solar Nebula \citep{Hay81}. Initially the disk contains 2000 planetesimals and 25 planetary embryos. The semi-major axes, masses and numbers of planetary embryos are chosen according to the concept of the isolation mass, which is the mass, that is contained in the planetary embryo's feeding zone. As a consequence, the initial separation of the planetary embryos is large enough, that they cannot collide immediately. Each planetesimal has a mass of $0.00118\ M_{\earth}$ and the masses of the planetary embryos range from $0.0582$ to $0.152\ M_{\earth}$. The disk has in total a mass of $M_{d,tot} = 4.8\ M_{\earth}$, half of which is contained in the planetesimals and half in the embryos. The fraction of the planetesimal mass of the total mass is larger than in similar studies \citep[see e.g.][]{KokIda02,Rayetal06}. The reason is an investigation by \citet{Gyeetal14} who studied a $\gamma$ Cephei like system using combined hydrodynamic - N-body simulations. They showed that in tight binary stars a larger fraction of non-accreting collisions than accreting ones can be expected. \\
Usually, the initial eccentricities and inclinations of disk objects following a Rayleigh distribution \citep{IdaMak92a} with dispersions $<e^2>^{1/2} = 2\cdot <i^2>^{1/2}$ which are assumed to be small.
However, \citet{Maretal09} pointed out, that in the preceding formation phase, planetesimals decoupled from the gas due to the forced inclination of the secondary star. Consequently, a planetesimal disk may evolve in a gas free environment in case of a misalignment between the secondary star and the disk larger than $10{\degr}$. In addition, test simulations have shown, that the inclination of the disk objects aligns with $i_{\mathrm{forced}}$ of the secondary star and exhibits variations of different $i_{\mathrm{free}}$, which also has been shown by \citet{Zhaetal18}. We therefore assume that an embryo-planetesimal disk in a misaligned binary star may not be dynamically cold after the gas phase\footnote{We refer here to the gas phase as the intermediate phase, i.e.~the phase when planetesimals grow to planetary embryos/cores and the gas has still a significant influence on the evolution of the planetesimals.}, i.e.~when our N-body simulations start. However, \citet{Xieetal11} argue, that when planetesimals traverse the gas disk at high inclinations, the gas drag they experience is amplified, due to the increased relative velocity between planetesimal and the gas disk, which would result in a stronger damping of the eccentricity and inclination of the planetesimal. Furthermore, when the inclinations of the disk objects relative to the secondary star's plane are high, they may undergo Kozai oscillations \citep{Koz62}, which cause them to spend a significant part of their time in low inclination but highly excited orbits. Since we do not consider highly misaligned configurations in this work (only $i_b=0\degr$ and $20\degr$), it is based on the work of \citet{Maretal09}. In principle, the simulations in inclined binary stars were started with dynamically excited disks (i.e.~larger values for the eccentricity and inclination of the disk objects). As the distribution of embryos and planetesimals in inclined binary stars after the gas phase is not known so far, we performed for some selected binary stars also calculations starting with planar disks (i.e.~initially dynamically cold disks with small values for eccentricity and inclination) in order to compare the results. The different initial states of the disks are indicated by the suffixes EIC (excited initial conditions) and PIC (planar initial conditions). Furthermore, in this work, the orbital elements are always given to the initial orbital plane of the embryo-planetesimal disk.\\ 
We did not include any gas giant in our simulations, and after the gas phase (or intermediate phase), a bi-modal distribution of planetary embryos and planetesimals remains \citep{KokIda00}. Therefore, the only external perturbing body on the disk objects is the secondary star (we refer here to systems like the Kepler 1229, 296 or 1620 system, which all inhabit rocky planets, but no giant planet).\\
General relativity (GR) is also irrelevant. The perihelion precession per revolution ($\delta\omega$) caused by GR can be approximated as follows:
\begin{equation}
  \delta\omega \approx \frac{6\pi G\left(M+m\right)}{a c^2 \left(1-e^2\right)}
\end{equation}
with $G$ the gravitational constant, $M$ the mass of the primary star, $c$ the speed of light, $m$, $a$, and $e$ the mass, semi-major axis and eccentricity of the planet, respectively. For the area of the disk in this work, the precession time is $\sim 3.3\cdot 10^7\ \mathrm{yr}$ for a body at $a=1\ \mathrm{au}$ and $1.29 \cdot 10^8\ \mathrm{yr}$ for a planet at $a=4\ \mathrm{au}$. The time scales are equivalent for a secondary star at $a_b\approx 550\ \mathrm{au}$ in the first case and $a_b\approx 1700\ \mathrm{au}$ in the latter case. As a consequence, the effects of GR are negligible for close binary star separations ($a_b\leq 100\ \mathrm{au}$) and the chosen simulation time in this work.

\subsection{Analysis of the disk}\label{sec:analysisDisk}
To analyze the dynamical behaviour of the disk objects, we computed mass weighted elements of the semi-major axis ($\langle a\rangle $), eccentricity ($\langle e\rangle $), inclination ($\langle i\rangle $), and argument of ascending node ($\langle \Omega\rangle $), which are defined as follows:
\begin{equation}
  \langle x\rangle  = \frac{\sum_j m_j\cdot x_j}{\sum_j m_j}
  \label{eq:mw}
\end{equation}
where $m_j$ are the masses of the disk objects, and $x_j$ the corresponding elements.\\
In addition, we applied the following quantities introduced by \citet{Cha01} and \citet{Quietal02} to describe the state of the computed planetary systems in order to compare them. \\
(i) A mass-concentration statistic is given by:
\begin{equation}
  S_c = \mathrm{max}\left\{\frac{\sum_j m_j}{\sum_j m_j \left[\log_{10}\left(a/a_j\right)\right]^2}\right\}
  \label{eq:massConc}
\end{equation}
where $m_j$ and $a_j$ are the mass and semi-major axis of each formed terrestrial planet. $S_c$ is given by the maximum value of the function in brackets as a function of $a$.\\
(ii) The radial-mixing statistic $S_r$ is given by:
\begin{equation}
  S_r = \left(\sum_j\frac{m_j\left|a_{\mathrm{init},j}-a_{\mathrm{fin},j}\right|}{a_{\mathrm{fin},j}}\right)/\sum_j m_j
\end{equation}
where $a_{\mathrm{init},j}$ and $m_j$ are the initial semi-major axes and the masses of the embryos successfully growing to terrestrial planets, and $a_{\mathrm{fin},j}$ are the final semi-major axes. $S_r$ gives a degree of the radial mixing which will be compared. For details of $S_c$ and $S_r$ see \citet{Cha01}.\\
(iii) The angular momentum (per unit mass) of the final $N$ planets is given by:
\begin{equation}
  l = \frac{\sum_j L_j}{\sum_j m_j} = \frac{\sum_j m_j\left[GM_{\sun} a_j\left(1-e_j^2\right) \right]^{1/2}}{\sum_j m_j}
\end{equation}
with $m_j$, $a_j$, and $e_j$ the planets masses, semi-major axes, and eccentricities of the formed planets, respectively. The total angular momentum (per unit mass) is normalized by  the initial total angular momentum (per unit mass) $l_0$.\\
(iv) The z-component of the angular momentum vector (per unit mass) $l_z$ is given by:
\begin{equation}
  l_z = \frac{\sum_j L_j \cdot\cos i_j}{\sum_j m_j}
\end{equation}
where $i_j$ is the planets' inclination with respect to the equator of the primary. The value is normalized by the initial z-component $l_{z,0}$ of the system. For details see \citet{Quietal02}.

\subsection{Quick stability check of the planetary systems after $10\ \mathrm{Myrs}$}
\subsubsection{Hill criterion}
The Hill criterion has often been used for a quick check of the stability of planetary systems \citep[see][for compact planetary systems]{HayTre98,Funetal10}. A system of two planets with masses $m_i$ and $m_{i+1}$ is Hill stable if the following inequality applies:
\begin{equation}
  a_{i+1} - a_i > k \cdot \left(R_{h,i} + R_{h,i+1}\right).
  \label{eq:hillStab}
\end{equation}
The planets move in circular orbits at distances $a_i$ (inner planet) and $a_{i+1}$ (outer planet) around a central star of mass $M$, $R_{h,i}$ and $R_{h,i+1}$ are the corresponding Hill radii, and $k$ is the multiple of Hill radii. However, since planets are not necessarily on circular orbits, but likely in elliptical motion, we used the aphelion distance instead of $a_i$ to calculate the Hill radius of the inner planet and the perihelion distance instead of $a_{i+1}$ for the outer planet. The equation of the Hill radius for the aphelion distance changes to
\begin{equation}
  R_{h,i} = a_i \cdot \left(1+e_i\right) \cdot \sqrt[3]{\frac{m_i}{3M}}
  \label{eq:hillApo}
\end{equation}
and for the perihelion distance to
\begin{equation}
  R_{h,i+1} = a_{i+1} \cdot \left(1-e_{i+1}\right) \cdot \sqrt[3]{\frac{m_{i+1}}{3M}}
  \label{eq:hillPer}
\end{equation} where $e_i$ and $e_{i+1}$ denote the eccentricity of the inner planet and outer planet, respectively.\\
In equation~(\ref{eq:hillStab}) we used $k=6.5$, which results in a distance of 13 Hill radii between to neighbouring planets \citep{Cha96}. Depending on the system, the critical distance for the Hill criterion is typically between $4$ and $15$ Hill radii. For details see \citet{SmiLiss09}.\\
However, since the Hill criterion (equation~(\ref{eq:hillStab})) is only an approximate estimate, we examined in addition the variations of the eccentricities of the planets and their changes. Our stability check resulted in the following classifications:\\
(i) {\it{stable}}--for all systems that satisfied the Hill criterion and where the planets had small variations in eccentricities,\\
(ii) {\it{semi-stable}}--for all systems that also satisfied the Hill criterion but where the planets had large variations ($\Delta e>0.1$) in eccentricity, and\\
(iii) {\it{unstable}}--for all systems that did not satisfy the Hill criterion.\\
The results for all binary configurations are summarized in Table~\ref{tab:binConfTP}.

\subsubsection{NAMD stability}
The NAMD (normalized angular momentum deficit) was introduced by \citet{Cha01} and gives the relative deficit of angular momentum, which is a further development of the AMD (angular momentum deficit) introduced by \citet{Las00} and further developed by \citet{LasPet17}. The AMD represents the violence of the dynamical evolution of a planetary system. It is defined as the difference between circular angular momentum (CAM), which is the norm of the angular momentum of the idealized planetary system with the same masses and semi-major axis values for the planets, but with circular and planar orbits and the norm of the angular momentum of the actual plane. The total AMD of a planetary system is given by
\begin{equation}
  \mathrm{AMD} = \sum_k m_k\sqrt{Gm_0a_k}\left(1-\sqrt{1-e^2_k}\cos i_k\right)
  \label{eq:AMD}
\end{equation}
with $m_k,a_k,e_k,i_k$ the planets mass, semi-major axis, eccentricity and orbital inclination respectively, and $m_0$ the stellar mass. \citet{Turetal20} suggest using the normalized angular momentum deficit (NAMD) by \citet{Cha01}, as it is better for comparison among different planetary systems. It is defined as the ratio between the AMD and the CAM:
\begin{align}
  \mathrm{NAMD} = S_d &= \frac{\mathrm{AMD}}{\mathrm{CAM}} \\
  &= \frac{\sum_k m_k \sqrt{a_k} \left(1-\sqrt{1-e^2_k}\cos i_k\right)}{\sum_k m_k\sqrt{a_k}}
  \label{eq:NAMD}
\end{align}
which is independent of the gravitational constant $G$ and the stellar mass $m_0$. Note, $i_k$ is measured with respect to the binary plane. The results for all binary configurations are summarized in Table~\ref{tab:binConfTP}.

\begin{figure*}
  \centering
  \includegraphics[width=\hsize]{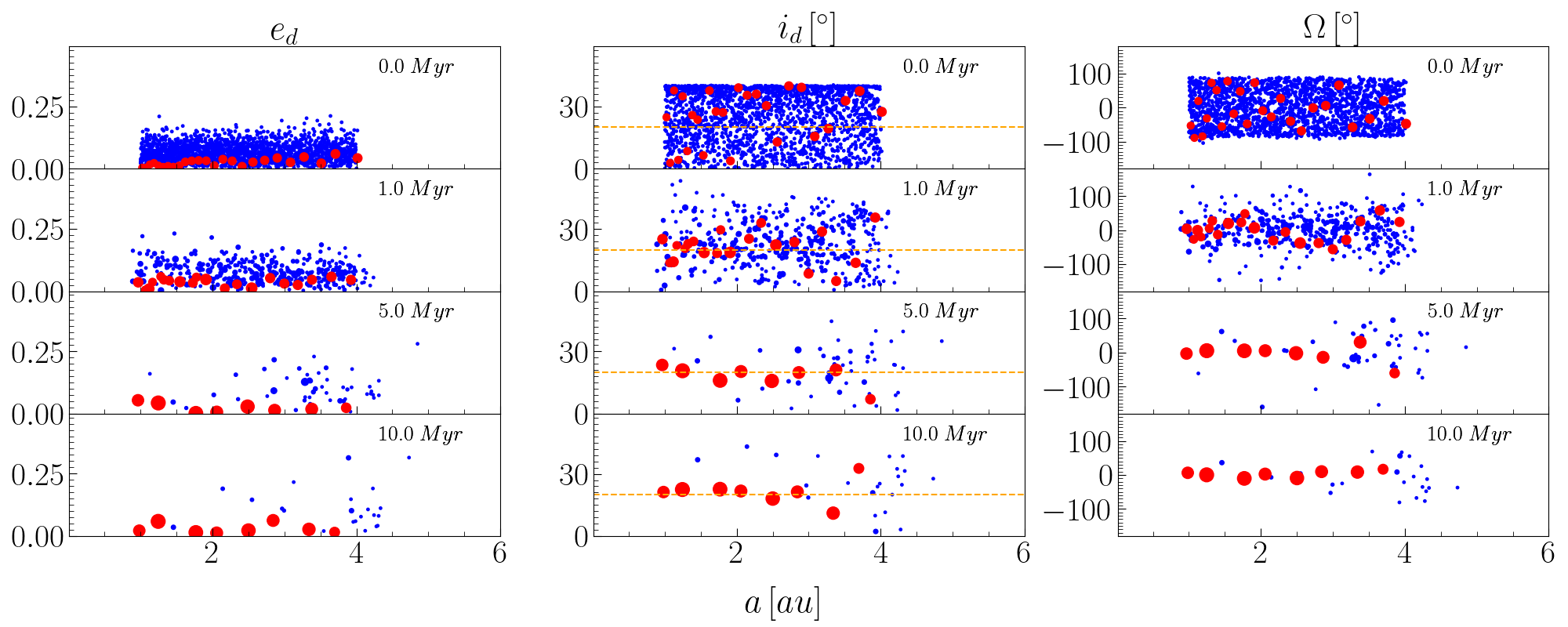}
  \caption{Snapshots of the evolution of eccentricity $e_d$, inclination $i_d$, and the longitude of the ascending nodes $\Omega_d$ of the disk objects for the configuration a30-e02-i20-EIC. The blue circles represent the planetesimals and the red ones the planetary embryos. The radius of the circles are proportional to their size. The orange dashed line in the middle panel indicates the inclination of the secondary star ($i_b=20\degr$).}
  \label{fig:multiTStep30a02e20iE}
\end{figure*}

\section{Results: Dynamical evolution of the disk}

\subsection{Effects of the binary star inclination (EIC configurations)}

\begin{figure*}
  \centering
  \includegraphics[width=\hsize]{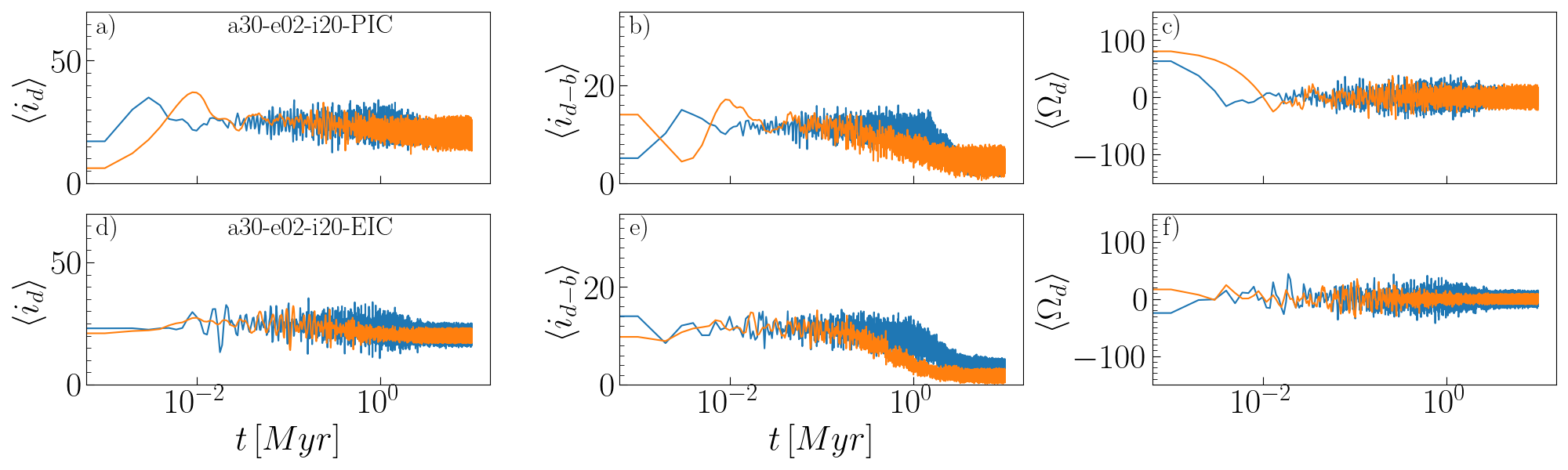}
  \caption{Comparison of the evolution of the inner disk of planetary embryos ($a_{\mathrm{emb}}\leq 2\ \mathrm{au}$, shown in orange) and the outer one ($a_{\mathrm{emb}}>2\ \mathrm{au}$, shown in blue) for the same configurations shown in Fig.~\ref{fig:semMaj}. Left panels show for each configuration the mass-weighted evolution of the inclination, middle panels the evolution of the mass-weighted inclination of the planetary embryos with respect to the binary stars plane $i_{d-b}\left(=\|i_d - i_b\| \right)$, and right panels the evolution of the mass-weighted ascending node.}
  \label{fig:iidbOM30}
\end{figure*}

\begin{figure*} 
  \centering
  \includegraphics[width=\hsize]{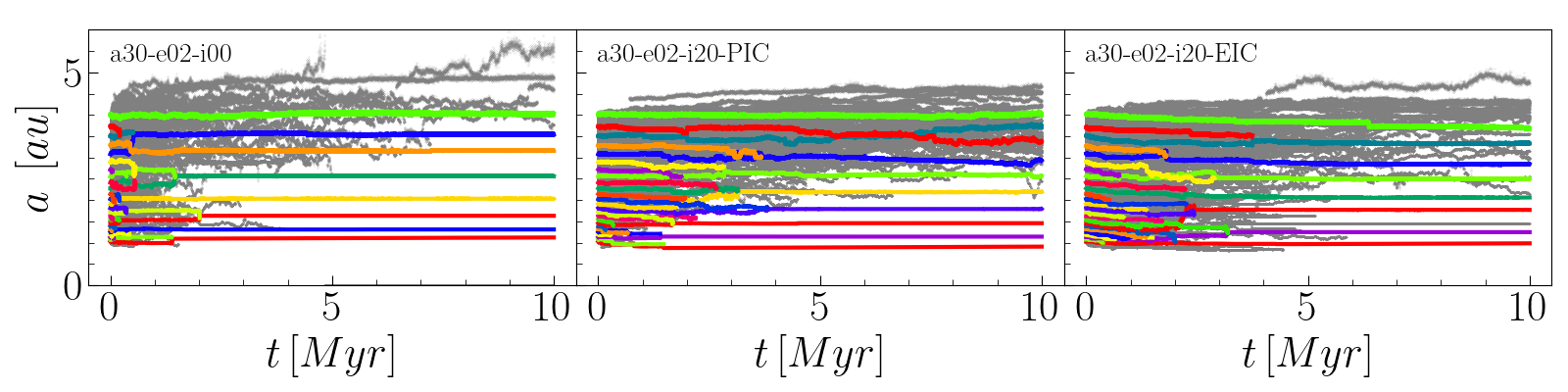}
    \includegraphics[width=\hsize]{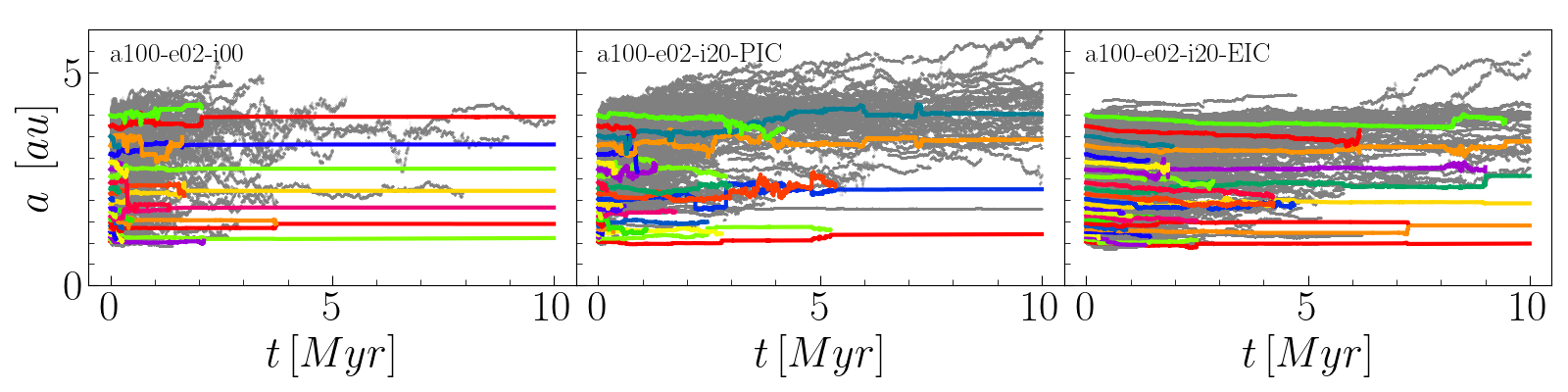}
  \caption{The evolution of the semi-major axis of the planetary embryos (color coded) and the planetesimals (gray) for $10\ \mathrm{Myr}$. In the top panels tight binary star configurations ($a_b=30\ \mathrm{au}$) are shown, while in the bottom panels the evolution in wider binary star configurations ($a_b=100\ \mathrm{au}$) is presented. The evolution is displayed for planar binary star configurations (left panels), inclined binary star configurations ($i_b=20\degr$) with an initially dynamically cold disk (middle panels), and inclined binary star configurations with an initially dynamically excited disk (right panels).}
  \label{fig:semMaj}
\end{figure*}

\begin{figure}
  \centering
  \includegraphics[width=\hsize]{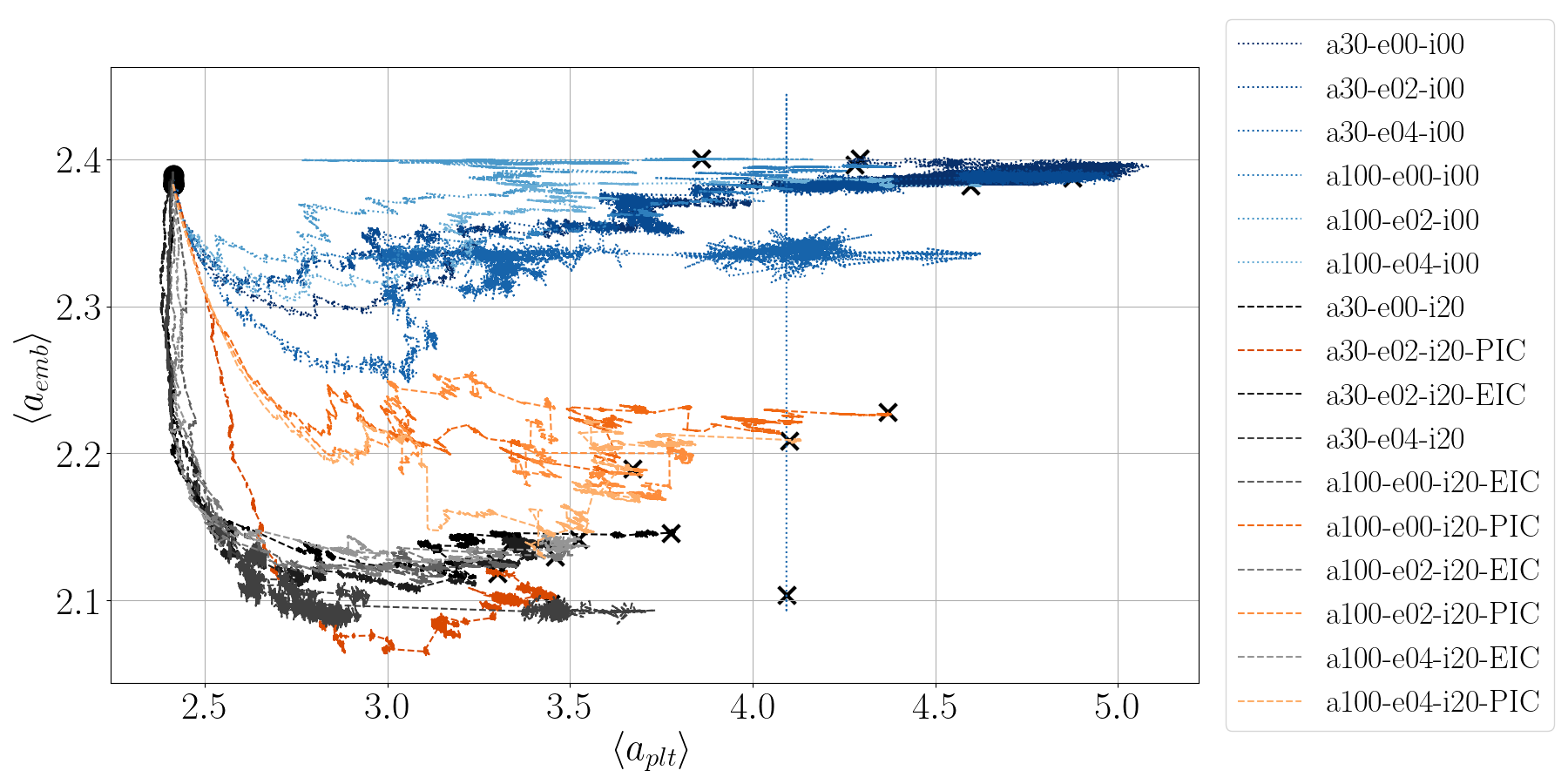}
  \caption{Overview of the disk objects' migration according to the evolution of the mass weighted semi-major axis (equation~(\ref{eq:mw})) of the planetary embryos $\langle a_{\mathrm{emb}}\rangle $ plotted against the mass weighted semi-major axis of the planetesimals $\langle a_{\mathrm{plt}}\rangle $ for all binary star configurations. The black circle shows the starting value at the beginning of the simulation, the black cross at the end of the simulation. Different line styles (dotted, dashed, and solid) correspond to different inclinations of the binary system $i_b$. Different colours correspond to different binary and initial disk configurations.}
\label{fig:MWA}
\end{figure}

Fig.~\ref{fig:multiTStep30a02e20iE} presents snapshots of the disk evolution in eccentricity $e_d$, inclination $i_d$ and ascending node $\Omega_d$. The middle panel shows clearly that the secondary star forces the inclination of the disk objects to align with $i_b$ (i.e.~the orange dashed line). \citep[This behavior has also been shown by][]{Maretal09,Zhaetal18}. In addition, the disk objects oscillate with $i_\mathrm{free}$ around $i_\mathrm{forced}$ (note that $i_{\mathrm{forced}}\approx i_b$). Due to damping effects, like dynamical friction \citep{Obretal06} or collisions, $i_\mathrm{free}$ can decrease, so that only small variations around $i_\mathrm{forced}$ occur. Fig.~\ref{fig:iidbOM30} shows the evolution of the mass-weighted inclination $\langle i_d\rangle$ (left panels), the mass-weighted inclination with respect to the binary plane $\langle i_{d-b}\rangle$ (middle panels) and the mass-weighted ascending node $\langle \Omega_d\rangle$ (right panels) of the planetary embryos in the inner part of the disk ($a_d \leq 2\ \mathrm{au}$, orange lines) and the outer part ($a_d > 2\ \mathrm{au}$, blue lines) for two different initial conditions in the misaligned a30-e02-i20 configurations. It can be seen, that damping occurs on a shorter time-scale in the inner part of the disk, due to the higher particle density and thus a more efficient damping (bottom middle panel of Fig.~\ref{fig:iidbOM30}). A similar evolution can be observed for the ascending node, which shows an alignment around the $\Omega_b$ with variations of $\Omega_{free}$. However, due to the aforementioned damping effects, their variations become small.\\
Fig.~\ref{fig:semMaj} shows the evolution in the semi-major axis of the disk objects for different binary star configurations and different initial disk set-ups. In the planar configurations (left panels), the evolution seems to be quite similar in the two systems. Most embryo-embryo collisions occur within the first $2\ \mathrm{Myr}$, which can be considered as the most chaotic phase of these systems. Collisions with planetesimals (gray lines) occur throughout the entire computation time. The planetary embryos do not show any migration effects and remain on their orbit after the chaotic phase. This is also reflected by the evolution of $\langle a_{emb}\rangle$ in Fig.~\ref{fig:MWA} for the planar configurations (blueish lines). They show no significant change in the y-axis, with one exception: System a30-e04-i00 shows a rapid decline in $\langle a_{emb}\rangle$ from $\sim 2.3$ to $\sim 2.1$ due to an ejection of an embryo.\\
However, the evolution of the semi-major axis of the disk objects in the inclined configurations differs from that in the planar configurations. In the EIC disks (right panels of Fig.~\ref{fig:semMaj}) the planetary embryos show a slight inward migration, which is visible especially during the first $1.5\ \mathrm{Myr}$. This behaviour might be similar to the migration of planetesimals in misaligned binary star systems described by \citet{Xieetal11}, which is caused by gas friction, where $da/dt \propto i^2$. Thus, larger misalignment indicate faster migration. In our study, however, the dissipative effects are due to dynamical friction and collisions. Migration stops when there are no other nearby bodies with which the planetary embryos can interact. Therefore, the migration process in the outer part of the disk can take place over a longer period of time, as the planetesimals are more likely to survive in the outer part of the disk. However, it is still possible, that the inward migration is turned around by mutual scattering of the embryos like the outer green embryo in the bottom right panel of Fig.~\ref{fig:semMaj}.

\subsection{Different initial disk set-ups (EIC vs.~PIC) in misaligned binary star configurations}\label{subsec:PIC}
\begin{figure*}
  \centering
  \includegraphics[width=\hsize]{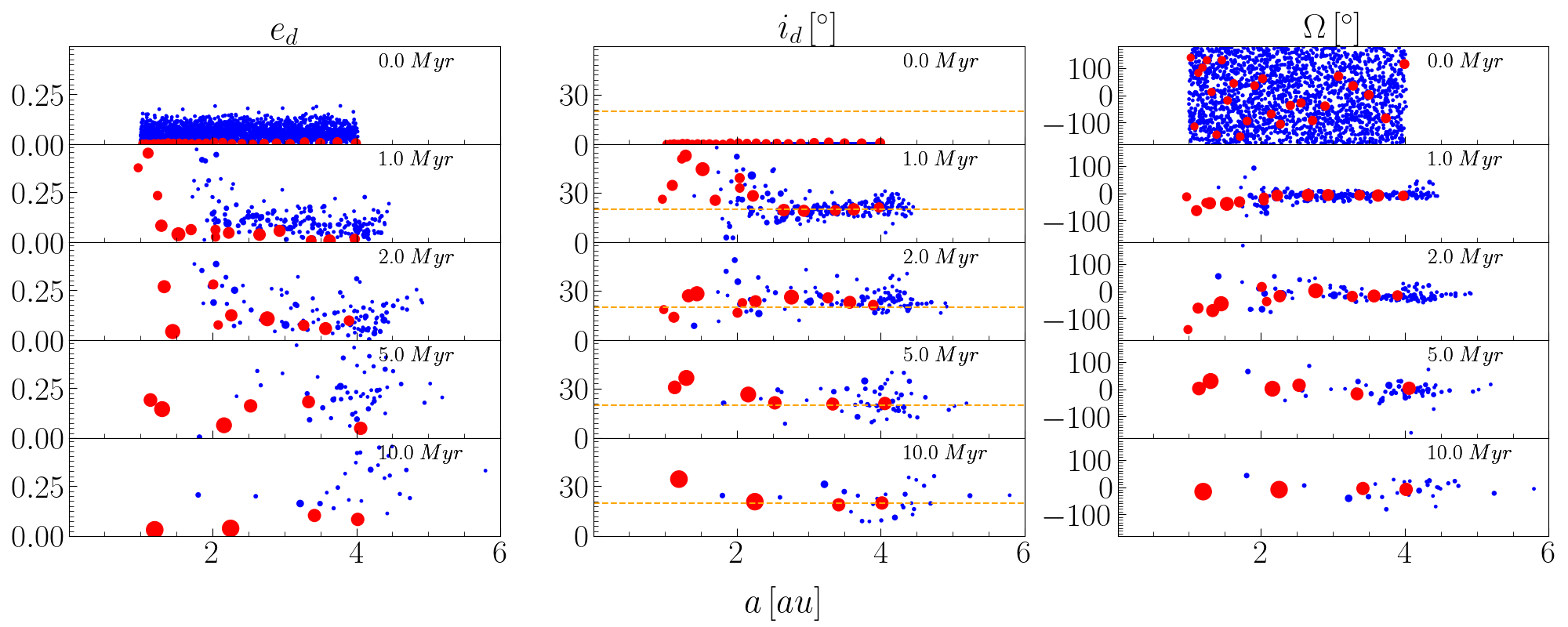}
  \caption{The same as in Fig.~\ref{fig:multiTStep30a02e20iE}, but for a initially dynamically cold planetary embryo-planetesimal disk and for the configuration: a100-e02-i20.}
  \label{fig:multiTStep100a02e20iP}
\end{figure*}

\begin{figure*}
  \centering
  \includegraphics[width=\hsize]{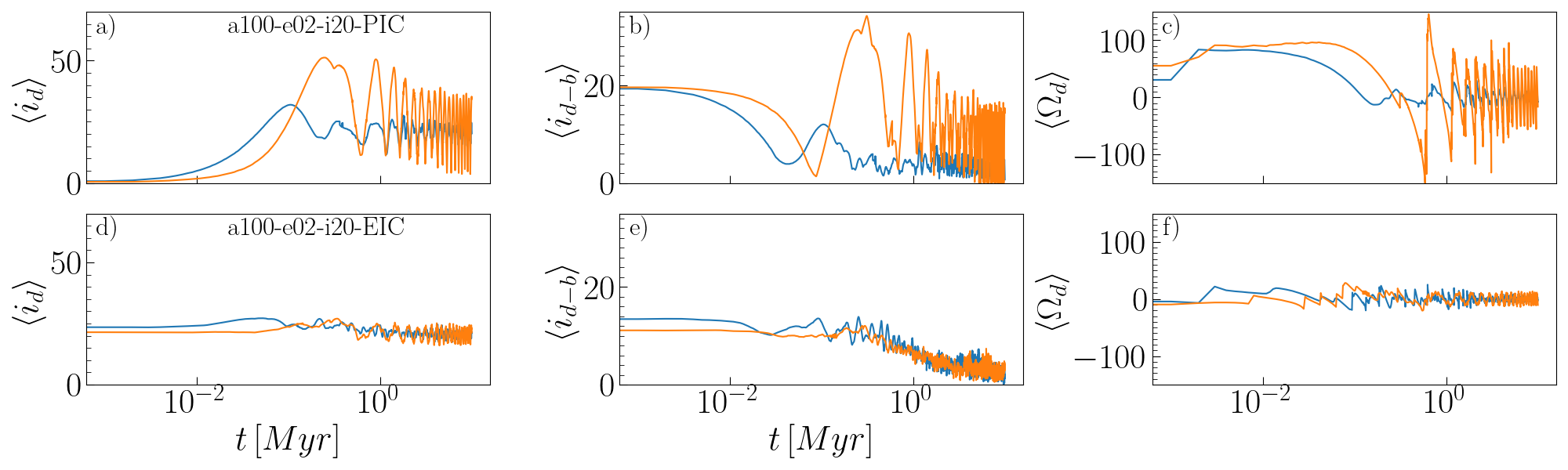}
  \caption{The same as Fig.~\ref{fig:iidbOM30} but for the configuration a100-e02-i20.}
  \label{fig:iidbOM100}
\end{figure*}

As mentioned in section~\ref{initCond}, the distribution of the disk objects after the intermediate phase (gas phase) is not yet fully understood, since both the interaction between the excitation of the secondary star and damping effects of the gas disk, as well as the parameters of the gas disk are somewhat unclear. Thus, we performed for selected inclined binary stars (see Table~\ref{tab:binConfTP}) also simulations starting with planar embryo-planetesimals disks which are indicated as PIC configurations while the according excited initial disks are marked as EIC configurations. A comparison of these systems shows that gravitational focusing and the stellar parameters (mainly the binaries' separation) play a crucial role in this context. Gravitational focusing is more effective in dynamically cold disks at the beginning of the simulation because the small eccentricities and inclinations that are quite similar for all disk objects. However, at some point in time the influence of the secondary star becomes noticeable, which depends on the stellar parameters. The time-scale of the influence can be approximated by the forced secular frequency induced by the secondary star which can be computed by the Heppenheimer model \citep{Hep78}. For a test particle this secular frequency is given by:
\begin{equation}
  g_P = \frac{3}{4} \left(\frac{m_B}{m_A}\right) \left(\frac{a_P}{a_B}\right)^3 n_P \left(1-e^2_B\right)^{-3/2},
\end{equation}
where $m_A$ and $m_B$ are the masses of the two binary stars, $a_B$ is the binary separation, $e_B$ the eccentricity, $a_P$ the semi-major axis of the particle, and $n_P$ the mean motion of the particle. Consequently, $1/g_P$ gives the corresponding period. The calculation for the configurations a30-e02-i20 and a100-e02-i20 yields secular periods between $5389-673\ \mathrm{yr}$ in the semi-major axis range from $1$ to $4\ \mathrm{au}$ for the tight binary star and between $199606 - 24950\ \mathrm{yr}$ for the wide binary star configuration. Thus, the alignment of the disk objects' inclination $i_d$ to $i_b$ occurs on longer time-scales for wider binary separations ($a_b = 100\ \mathrm{au}$) than for tight binary configurations ($a_b=30\ \mathrm{au}$), which can be seen when comparing the first peaks (orange and blue curves) in the top left panels of Figs~\ref{fig:iidbOM30} and~\ref{fig:iidbOM100}. Thus, gravitational focusing (due to the lower inclinations of the disk objects) is more effective on a longer time-scale in the wider binary configurations, which increases the collision efficiency. By the time the secular perturbations affect the inner region of the disk, most planetesimals have already collided. As a result the damping due to dynamical friction and collisions is less effective, and cannot reduce the eccentricity and the $i_{free}$ of the inner disk objects (indicated in the left and middle panels of Fig.~\ref{fig:multiTStep100a02e20iP}). This leads to a completely different evolution for the inner disk in the case of a PIC configuration (see Fig.~\ref{fig:iidbOM100} upper panels) than for the EIC configuration (Fig.~\ref{fig:iidbOM100} lower panels). Since no damping occurs through the interactions with the planetesimals, the inner embryos can reach large eccentricities and inclinations (see Fig.~\ref{fig:multiTStep100a02e20iP}). Only mutual interactions of the remaining embryos cause a slight damping of inclination, eccentricity, and node. Furthermore, inward migration is less effective, as shown by the evolution of $\langle a_{emb}\rangle$ in Fig.~\ref{fig:MWA}.\\
However, in the tight binary configuration, the secular perturbations act on a shorter time-scale. Therefore, the entire disk is excited shortly after the start of the simulation, so that gravitational scattering can only occur at this time. Due to this short time-scale of the forced secular frequency, the evolution of an initially dynamically cold disk (PIC system) does not differ much from the EIC configuration. The inner part of the disk shows slightly larger variations in $i_d$ and $\Omega_d$ (orange curves in the top panels of Fig.~\ref{fig:iidbOM30}).

\subsection{Architectures after 10 Myrs}

The simulations of the various systems naturally show different configurations after $10$ million years of calculation. Table~\ref{tab:binConfTP} gives an overview of the architectures of all systems, the first part shows the number of meanwhile formed terrestrial planets (column 2), their mass range (column 3), the area of their final locations (column 4), their eccentricities (column 5) and their inclinations (column 6). The second part of Table~\ref{tab:binConfTP} provides information about the stability of the systems after $10\ \mathrm{Myrs}$ (columns 7 and 8) and further statistical values, described in section~\ref{sec:analysisDisk}: $S_c$ (column 9), $S_r$ (column 10), $l/l_0$ (column 11), and $l_z/l_{z,0}$ (column 12). Figs~\ref{fig:finConfPlan},~\ref{fig:finConf20}, and~\ref{fig:finConfPIC} in the appendix show the architectures after $10\ \mathrm{Myrs}$ simulation time of the planar, the i20-EIC, and the i20-PIC configurations, respectively. \\
Most terrestrial planets are formed within the planar binary configurations. On average about 8 bodies (column 2 in Table~\ref{tab:binConfTP}) with masses between $0.18 - 1.38\ M_{\earth}$ were formed. The configuration a30-e04-i00 is an outlier, with the largest planet being $2.12\ M_{\earth}$ and due to an ejection of one planet, only 4 terrestrial planets remain. In the i20-EIC configurations slightly less terrestrial planets have been formed, on average 6-7. They have masses between $0.18-1.22\ M_{\earth}$, which are a bit smaller compared to the planar configurations. In the i20-PIC configurations the formed planets depend strongly on the separation of the binary stars. In a30-e02-i20-PIC 10 bodies with masses between $0.162-0.846\ M_{\earth}$ have been formed, which are smaller than those in the i20-EIC configurations. The wider configurations (a100-PIC) host only 5-6 terrestrial planets, but with masses larger between $0.188 - 1.78\ M_{\earth}$. \\
The formed planets or proto-planets are distributed between $1.04 - 4.24\ \mathrm{au}$, indicating a small scattering outward of the initial disk boundaries for some binary configurations. Again, configuration a30-e04-i00 is an outlier, due to the ejection of a planet, the outermost planet is at $2.69\ \mathrm{au}$. In the i20-EIC configurations, small inward migration is recognizable, as most of the outer planets or proto-planets are within $3.5\ \mathrm{au}$. The semi-major axis range is between $0.96-3.94\ \mathrm{au}$. In the case of the i20-PIC configurations, the terrestrial planets show a slight scattering for all 4 configurations. However, the inner terrestrial planet in the a30-e02-i20-PIC shows an inward migration to $0.906\ \mathrm{au}$, which is not the case for the a100-PIC configurations. The eccentricity (column 5) of the formed objects are generally low in all configurations. The largest ones occur in the i20-PIC configurations with up to $0.137$. As in the planar configurations no misaligned perturber is present, the terrestrial planets have low inclinations. In the inclined configurations the range of the inclinations is distributed around $i_\mathrm{forced}=20\degr$, in both the i20-EIC and i20-PIC configurations. However, in the a100-PIC configurations the range is slightly larger.\\
For an estimate on the stability, the Hill criteria (equation.~(\ref{eq:hillStab})) and NAMD ($S_d$, equation.~(\ref{eq:NAMD})) stability have been computed and are presented in columns (7) and (8). Sometimes the two quantities yield different results for the same configuration. In the planar configurations, the Hill criteria estimates stable, semi-stable and unstable configurations. However, the NAMD stability yields values below $S_d<0.0018$, which is the value for the terrestrial planets of the solar system \citep{Quietal02} and what we denote as stable for the lifetime of the Sun. In the inclined configurations (i20-EIC), it is quite the opposite, as some of the configurations which are estimated as Hill stable have values $S_d > 0.0018$, thus larger values than in the planar configurations and can be denoted as less stable according to the NAMD stability. On the other hand, in the PIC configurations, a30-e02-i20 has the lowest value of $S_d = 0.0063$, but is estimated as Hill unstable, while the a100-PIC configurations all have larger values of $S_d$, but are either denoted as Hill stable or semi-stable. Of course, we are aware of the fact that the evolution of the systems are not finished after $10\ \mathrm{Myrs}$. As a consequence the different computations of the stability estimates yield different results. In general, determining the stability of a system a longer simulation time would be needed.\\
The statistical quantities provide the following information. The mass concentration statistic $S_c$ (equation~(\ref{eq:massConc})) gives an estimate of the extent to which the mass is concentrated in a part of the planetary system. In the case of the solar system with the four terrestrial planets $S_c=89.9$, since most of the mass is contained in the Earth and Venus. The here formed systems have significant smaller values ($S_c=4.655 - 9.398$). Due to the absence of a Jupiter-like planet (similar mass and semi-major axis), terrestrial planets can survive up to $4\ \mathrm{au}$ and thus the final mass is distributed between $1-4\ \mathrm{au}$. The largest value ($9.398$) is a result of an ejection of a terrestrial planet in configuration a30-e04-i00, which increases the value of the mass concentration.\\
The radial mixing statistic, $S_r$ is an estimate of each planet's radial migration relative to its initial position. Therefore, the value at the end of the simulation does not reflect its migration history. Due to the absolute value in the formula, it cannot be distinguished between inward and outward migration. A value of $S_r=0$ would indicate no migration. Larger values indicate a stronger migration behaviour. Interestingly, the planar binary configurations show larger values of $S_r=0.068 - 0.378$ than in the misaligned binary configurations ($S_r=0.022 - 0.0984$ and $S_r=0.05-0.0974$), which show a slight inward migration according to Figs~\ref{fig:semMaj} and~\ref{fig:MWA}.\\
The total angular momentum (per unit mass) increases slightly in the planar binary configurations, except in the configuration a30-e04-i00, due to the ejection of a planetary embryo, where it decreases about $5-6\%$. The z-component of the total angular momentum shows similar values. For the misaligned EIC configurations the total angular momentum decreases about $5-7\%$, while the z-component shows both an increase and decrease between $1-4\%$. On the other hand, in the PIC configurations $l$ decreases by $5-6\%$ and $l_z$ by up to $\sim 15\%$.

\begin{table*}
  \caption{Characteristics of the final planetary system for each binary configuration. $N_{TP}$ is the number formed terrestrial planets. The columns {\bf Mass-}, $\bm{a}$-, $\bm{e}$-, and $\bm{i}${\bf -range} indicate the maximum and minimum value of the formed terrestrial planets at the end of each simulation. {\bf Stability} represents the Hill stability of each system: $'+'$ denotes Hill stable, $\sim$ semi-stable, and $'-'$ unstable. The EIC configurations marked as bold are the ones also computed as PIC.}
  \label{tab:binConfTP}
  \centering
    \resizebox{\textwidth}{!}{\begin{tabular}{c || c l l l l c c c c c c}
    \hline
    \multicolumn{12}{c}{\bf Terrestrial planets}\\
    \hline
    {\bf Configuration} & $\bm{N}_{TP}$ & {\bf Mass-range} $\left[M_{\earth}\right]$ & $\bm a${\bf -range} $\left[\mathrm{au}\right]$ & $\bm{e}${\bf -range} & $\bm{i}${\bf -range} $\left[{\degr}\right]$ & {\bf Hill Stability} & $S_d$ & $S_c$ & $S_r$ & $l/l_0$ & $l_z/l_{z,0}$\\
    \hline
    \multicolumn{1}{c||}{(1)} & \multicolumn{1}{c}{(2)} & \multicolumn{1}{c}{(3)} & \multicolumn{1}{c}{(4)} & \multicolumn{1}{c}{(5)} & \multicolumn{1}{c}{(6)} & \multicolumn{1}{c}{(7)} & \multicolumn{1}{c}{(8)} & \multicolumn{1}{c}{(9)}& \multicolumn{1}{c}{(10)}& \multicolumn{1}{c}{(11)}& \multicolumn{1}{c}{(12)}\\
    \hline
    a30-e00-i00 & 8 & $0.26 - 1.14$ & $1.08 - 4.24$ & $0.015 - 0.11$ & $0.28 - 1.43$ & $\sim$ & $0.0016$ & $6.169$ & $0.1175$ &  $1.00162$ & $1.00158$ \\
    a30-e02-i00 & 8 & $0.26 - 1.1$ & $1.12 - 4.04$ & $0.004 - 0.05$ & $0.13 - 1.66$ & $+$  & $0.000417$ & $6.442$ & $0.0694$ & $1.00147$ & $1.00151$ \\
    a30-e04-i00 & 4 & $0.47 - 2.12$ & $1.17 - 2.69$ & $0.025 - 0.126$ & $1.54 - 2.55$ & $\sim$ & $0.0007$ & $9.398$ & $0.2243$ & $0.9416$ & $0.94121$ \\
    a60-e00-i00 & 8 & $0.26 - 1.16$ & $1.16 - 4.16$ & $0.02 - 0.049$ & $0.3 - 2.34$ & $\sim$ & $0.0008$ & $6.257$ & $0.0684$ & $1.00295$ & $1.00295$ \\
    a60-e02-i00 & 8 & $0.27 - 1.07$ & $1.09 - 4.01$ & $0.01 - 0.066$ & $0.22 - 2.03$ & $+$ & $0.00081$ & $6.195$ & $0.0738$ & $1.00278$ & $1.00273$\\
    a60-e04-i00 & 7 & $0.33 - 1.04$ & $1.13 - 4.09$ & $0.012 - 0.094$ & $0.05 - 0.6$ & $\sim$ & $0.0011$ & $6.282$ & $0.0727$ & $1.00223$ & $1.00228$\\
    a100-e00-i00 & 10 & $0.29 - 0.74$ & $1.07 - 4.08$ & $0.018 - 0.041$ & $0.24 - 0.71$ & $-$ & $0.000479$ & $6.271$ & $0.378$ & $1.00282$ & $1.00287$\\
    a100-e02-i00 & 7 & $0.48 - 0.86$ & $1.11 - 3.96$ & $0.005 - 0.041$ & $0.04 - 0.67$ & $+$ & $0.00048$ & $6.254$ & $0.095$ & $1.00359$ & $1.00362$\\
    a100-e04-i00 & 8 & $0.18 - 1.38$ & $1.04 - 3.54$ & $0.016 - 0.093$ & $0.13 - 1.66$ & $\sim$ & $0.000695$ & $6.384$ & $0.07$ & $1.00021$ & $1.00026$\\
    \hline
    a30-e00-i20-EIC & 8 & $0.18 - 0.95$ & $0.96 - 3.79$ & $0.009 - 0.1$ & $10.53 - 29.256$ & $\sim$ & $0.005$ & $6.162$ & $0.0482$ & $0.94818$ & $0.98451$ \\
    \textbf{a30-e02-i20-EIC} & 8 & $0.23 - 09$ & $0.98 - 3.7$ & $0.012 - 0.063$ & $11.08 - 32.63$ & $\sim$ & $0.0016$ & $6.23$ & $0.0795$ & $0.9479$ & $1.01381$ \\    
    a30-e04-i20-EIC & 5 & $0.7 - 1.22$ & $1.05 - 3.15$ & $0.01 - 0.066$ & $17.7 - 22.33$ & $+$ & $0.00133$ & $6.666$ & $0.022$ & $0.93971$ & $0.99341$ \\
    a60-e00-i20-EIC & 7 & $0.25 -0.91$ & $0.97 - 3.94$ & $0.013 - 0.053$ & $16.68 - 22.43$ & $\sim$ & $0.00148$ & $6.438$ & $0.0404$ & $0.94456$ & $1.02507$\\
    a60-e02-i20-EIC & 7 & $0.41 - 1.15$ & $1.03 - 3.72$ & $0.0187 - 0.052$ & $15.44 - 32.1$ & $+$ & $0.00448$ & $6.34$ & $0.37$ & $0.946412$ & $0.96051$\\
    a60-e04-i20-EIC & 6 & $0.59 - 1.09$ & $1.02 - 3.43$ & $0.02 - 0.074$ & $17.94 - 25.93$ & $+$ & $0.002$ & $6.298$ & $0.0636$ & $0.94404$ & $0.98043$\\
    \textbf{a100-e00-i20-EIC} & 5 & $0.7 - 1.05$ & $1.07 - 3.39$ & $0.047 - 0.061$ & $15.56 - 23.75$ & $+$ & $0.002772$ & $5.898$ & $0.0502$ & $0.9423$ & $0.992$ \\    
    \textbf{a100-e02-i20-EIC} & 5 & $0.61 - 1.11$ & $0.98 - 3.38$ & $0.04 -0.068$ & $16.92 - 25.41$ & $+$ & $0.003026$ & $6.06$ & $0.0508$ & $0.9439$ & $0.9803$\\
    \textbf{a100-e04-i20-EIC} & 6 & $0.56 - 1.03$ & $1.05 - 3.56$ & $0.036 - 0.11$ & $13.66 - 24.14$ & $\sim$ & $0.0049$ & $5.92$ & $0.0984$ & $0.94411$ & $1.00927$\\
    \hline
    a30-e02-i20-PIC & 10 & $0.162 - 0.846$ & $0.906 - 4.027$ & $0.009 - 0.137$ & $11.57 - 33.34$ & $-$ & $0.0063$ & $5.94$ & $0.05$ & $0.9402$ & $0.863$\\
    a100-e00-i20-PIC & 5 & $0.22 - 1.78$ & $1.3 - 4.11$ & $0.006 - 0.15$ & $16.56 - 40.61$ & $+$ & $0.0263$ & $5.893$ & $0.05$ & $0.96145$  & $0.8369$\\
    a100-e02-i20-PIC & 4 & $0.53 - 1.78$ & $1.2 - 4.02$ & $0.031 - 0.1$ & $18.86 - 34.33$ & $+$ & $0.011$ & $5.025$ & $0.0974$ & $0.9514$ & $0.8614$\\
    a100-e04-i20-PIC & 6 & $0.188 - 1.308$ & $1.058 -4.237$ & $0.0043-0.114$ & $4.26-27.71$ & $\sim$ & $0.0106$ & $4.655$ & $0.066$ & $0.95538$ & $0.90585$\\
  \end{tabular}}
\end{table*}

\section{Results: Collisions}
Terrestrial planets form by the growth of embryos due to collisions with other disk objects. Consequently, the collisions themselves must be considered in detail. In this context, we show the differences between planar and inclined binary configurations regarding the collision outcomes. Due to the low eccentricities and inclinations of the disk objects in a dynamically cold disk, the embryos grow faster than in an excited disk. In addition, the collisions of the disk objects are mostly accretive due to the low impact velocities. In a dynamically excited disk, on the other hand, the eccentricities and inclinations of the disk objects are larger and indicating a higher variation of $i_d$, resulting in less effective gravitational focusing and, consequently slower grow of the embryos. Furthermore, the high eccentricities and inclinations can lead to higher impact velocities, so that collisions may no longer be accretionary but rather destructive.\\
In addition, the embryos grow more slowly in the inclined configurations than in the planar configurations due to the lower collision probability caused by the lower disk density, and the weaker gravitational focusing. The proportion of embryo mass in planar configurations after $1\ \mathrm{Myr}$ is about $94\%$ of the total mass and after $10\ \mathrm{Myrs}$ $99\%$. In the inclined configurations (EIC), the total embryo mass after $1\ \mathrm{Myr}$ is about $72\%$. This percentage increases after $10\ \mathrm{Myrs}$ to about $99\%$ for the i20-EIC configurations. 

\subsection{Collision outcome maps}
\begin{figure}
  \centering
  \includegraphics[width=\hsize]{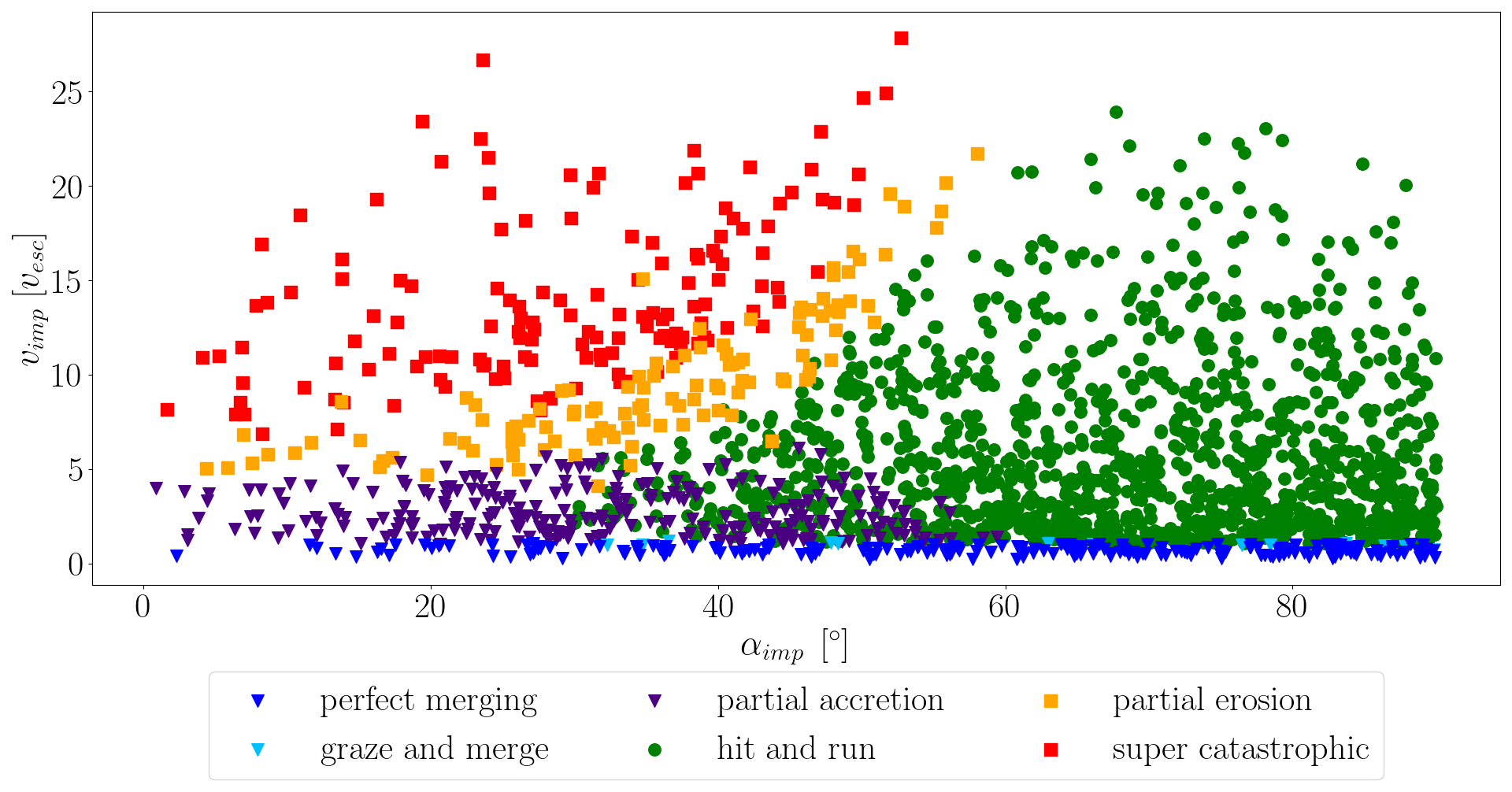}
  \caption{The different collision outcomes of all collisions for the binary star configuration: $a_b=30\ \mathrm{au}$, $e_b=0.2$, and $i_b=20{\degr}$, dynamically excited initial conditions (EIC). The x- and y-axis show the impact angle $\alpha$ and normalized impact velocity $v_{imp}$ for each collision. Colours and symbols indicate the collision outcome following \citep{LeiSte12}. The bluish colours indicate accretive collisions, the reddish ones destructive collisions and green circles represents hit-and-run collisions.}
  \label{fig:collOutcomeFull60a00e45i}
\end{figure}

\begin{figure}
  \centering
  \includegraphics[width=\hsize]{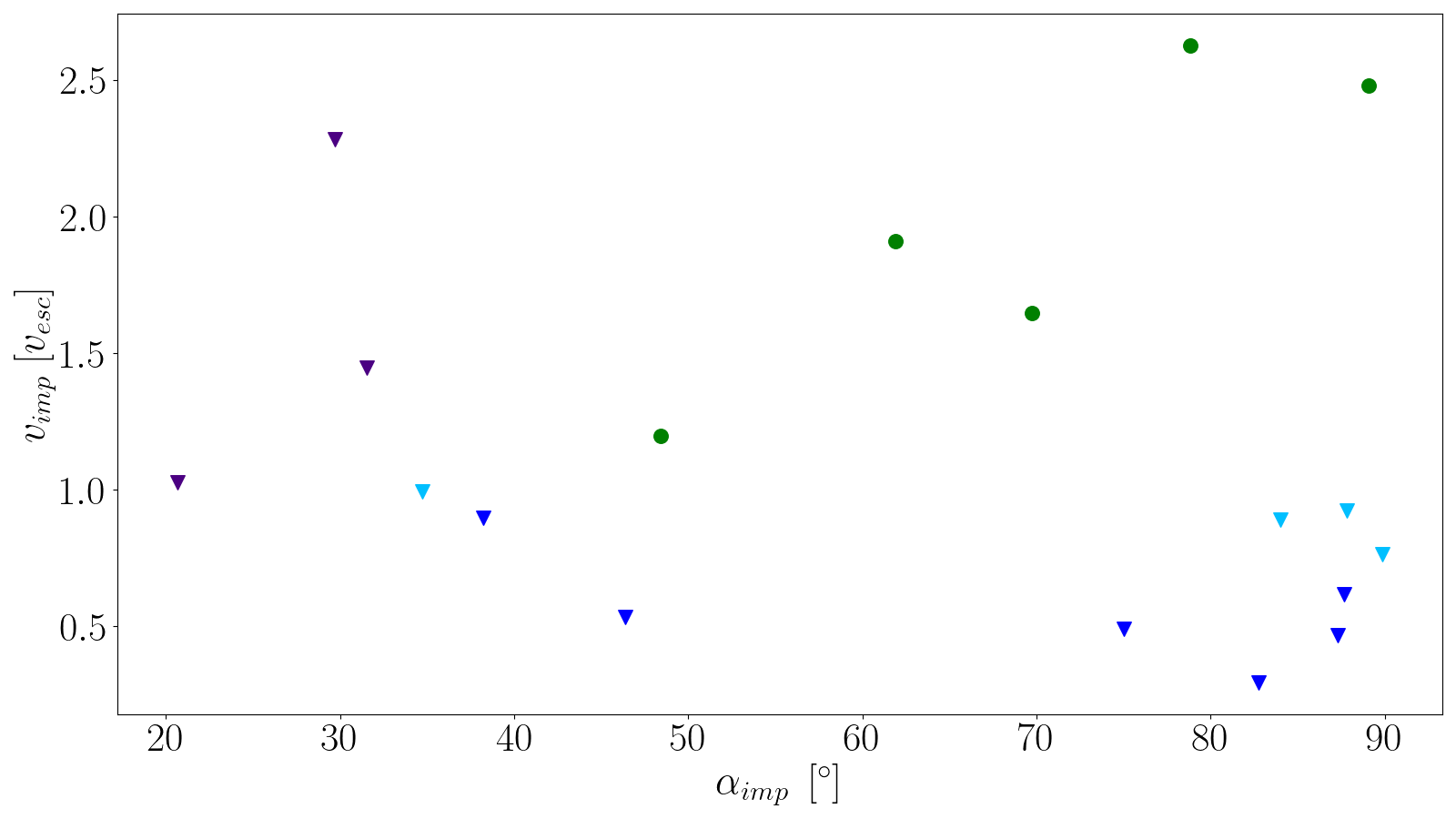}
  \includegraphics[width=\hsize]{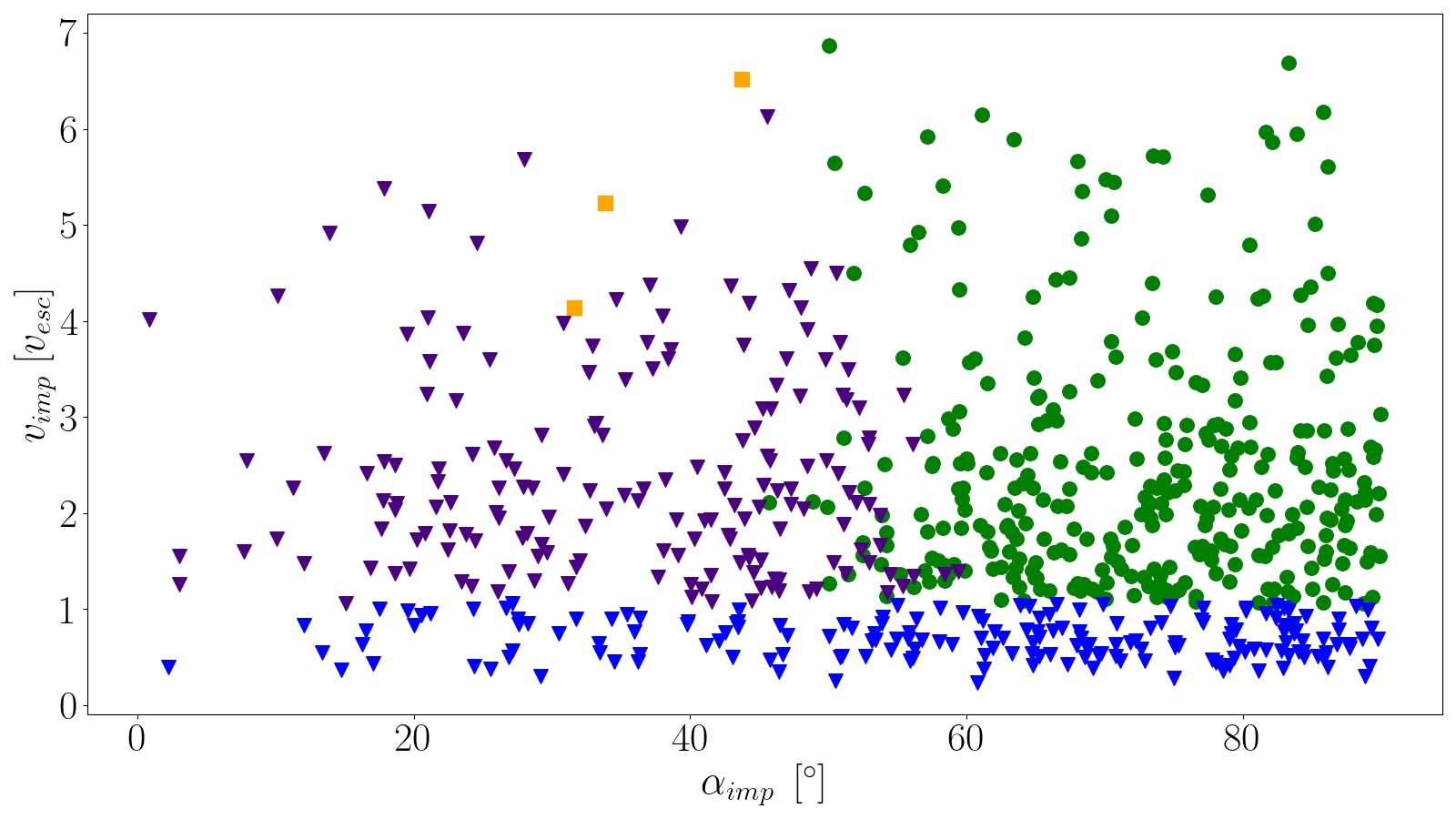}
  \includegraphics[width=\hsize]{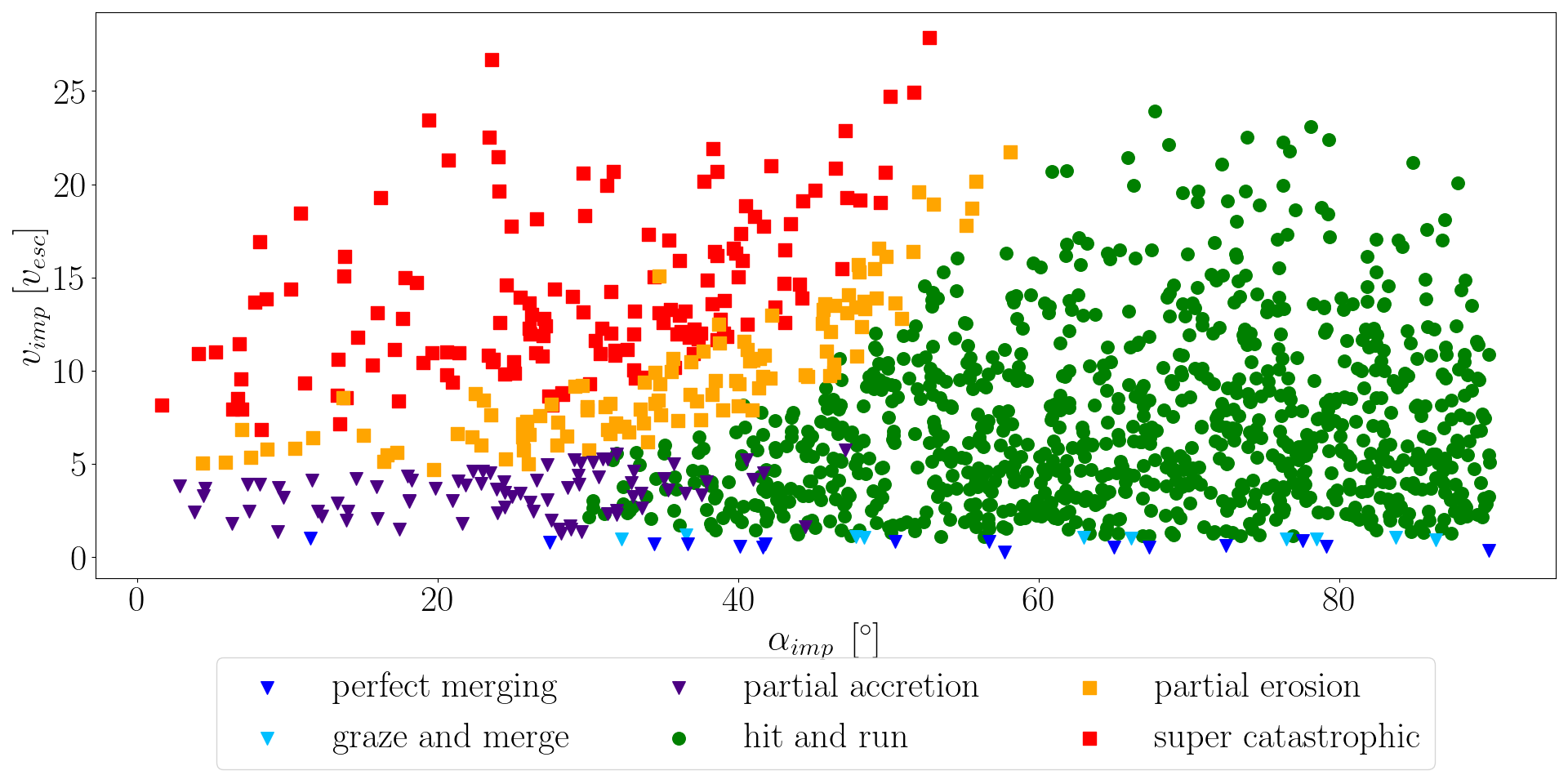}
  \caption{Same as Fig.~\ref{fig:collOutcomeFull60a00e45i} (configuration a30-e02-i20-EIC), but separated for the three different collision pairs. Top panel: embryo-embryo (e-e) collisions, middle panel: embryo-planetesimal (e-p) collisions, and bottom panel: planetesimal-planetesimal (p-p) collisions. Color code and symbols are the same as in Fig.~\ref{fig:collOutcomeFull60a00e45i}}
  \label{fig:collOutcomeSeperated60a00e45i}
\end{figure}

All collision that occurred in our N-body simulations were analysed in a post-processing step that used the analytical collision model of \citet{LeiSte12} to determine how a more accurate collision handling algorithm could influence the result. The model of (\citet{LeiSte12}) computes the collision outcome mainly based on (i) the impact velocity normalized on the colliding objects' escape velocities ($v_{imp}$), (ii) the impact angle ($\alpha$), and (iii) the mass ratio ($\gamma$) of the two colliding objects. Thus, in our study we tracked these parameters for each collision that occurred during the N-body simulations and applied the model of \citet{LeiSte12} for each collision using the values $c^\ast = 1.9$ and $\bar{\mu}=0.36$, where $c\ast$ is a scaling constant, which is equal to the offset of the gravitational binding energy, and $\bar{\mu}$ dimensionless material constant related to the energy and momentum coupling between projectile and target. Both constants are fitted by \citet{LeiSte12} from numerical studies and used in their work. We distinguished six different types of collision outcomes according to the study by \citet{LeiSte12}:
\begin{itemize}
\item {\it Perfect merging}: The two bodies merge perfectly
\item {\it Graze and merge}: The two bodies collide, separate with a small velocity and merge shortly afterwards in a second collision
\item {\it Partial accretion}: A part of the impacting body is accreted by the target
\item {\it Partial erosion}: The target body loses some material
\item {\it Super catastrophic}: Both bodies are destroyed during the collision. The largest remaining body has a mass $m_{lr} < m_{tot}$.
\item {\it Hit-and-run}: The bodies bounce off and are separated again
\end{itemize}
An example of such a post-processing analysis of the recorded collisions is shown in Fig.~\ref{fig:collOutcomeFull60a00e45i} for the configuration a30-e02-i20-EIC. This figure shows only 206 perfect merging events out of 2004 collisions and graze and merge events were even rarer, only 14 times. Partial accretion occurred more frequently (260) and most collisions (1263) were hit-and-run outcomes, followed by super-catastrophic events (145) and partial erosion (116). Due to the different mass ratios of the colliding bodies, we have found overlaps of the different collision outcome regions, e.g.~partial accretion and hit-and-run between $\alpha = 30 - 60{\degr}$. However, the picture looks different when splitting the result into the three different possible collision pairs: (i) embryo-embryo (e-e), (ii) embryo-planetesimal (e-p), and (iii) planetesimal-planetesimal (p-p) collisions. The top panel of Fig.~\ref{fig:collOutcomeSeperated60a00e45i} shows the outcomes for e-e collisions, which are by far the least occurring ones. It is clearly seen, that most of the collision outcomes are accretive ones (i.e.~perfect merging, graze and merge, and partial accretion), followed by hit-and-run collisions. No destructive collision, i.e.~partial erosion, and super catastrophic occurred. More common are e-p collisions (in total 686) shown in the middle panel of Fig.~\ref{fig:collOutcomeSeperated60a00e45i}. Here, the largest fraction of all collisions are hit-and-run outcomes (323), followed by the partial accretion (176) and perfect merging (184) events. Additionally, 52 partial erosive outcomes were found. Most collisions (in total 1300) occurred between planetesimals themselves (Fig.~\ref{fig:collOutcomeSeperated60a00e45i} bottom panel). The largest fraction is again the hit-and-run regime (935), followed by partial accretion (260) and perfect merging (206) collisions. More partial erosion (116) and super-catastrophic (145) collisions happened than for e-p pairs. Only a few graze and merge collisions (14) are visible.\\

\begin{table}
\caption{The table shows the total collisions summarized for both the planar and the misaligned (EIC) binary configurations. The total number of collisions is divided into the three possible collision pairs, and the corresponding fractions of the different collision results. $N_{\mathrm{coll}}$ are the respective numbers of collisions, $f_{\mathrm{accr}}$, $f_{\mathrm{destr}}$ and $f_{\mathrm{hr}}$ the respective fractions in accretive collisions (perfect merging, graze and merge and partial accretion), destructive collisions (partial erosion and super catastrophic) and hit-and-run collisions.}
\label{tab:bin_confCollSum}
\centering
\begin{tabular}{ccccc}
  &\multicolumn{4}{c}{\textbf{e-e}}\\
  Configurations & $N_\mathrm{coll}$ & $f_\mathrm{accr}$ & $f_\mathrm{destr}$ & $f_\mathrm{hr}$\\
  \hline
  planar & 156 & 100 & 0 & 0\\
  i20-EIC &173 & 62.43 & 0.58 & 36.99\\
  \hline
  \hline
  &\multicolumn{4}{c}{\textbf{e-p}}\\
  Configurations & $N_\mathrm{coll}$ & $f_\mathrm{accr}$ & $f_\mathrm{destr}$ & $f_\mathrm{hr}$\\
  \hline
  planar & 8336 & 99.18 & 0 & 0.82\\
  i20-EIC &6090 & 54.01 & 0.67 & 45.24\\
  \hline
  \hline
  &\multicolumn{4}{c}{\textbf{p-p}}\\
  Configurations & $N_\mathrm{coll}$ & $f_\mathrm{accr}$ & $f_\mathrm{destr}$ & $f_\mathrm{hr}$\\
  \hline
  planar & 9638 & 35.98 & 0.19 & 63.83\\
  i20-EIC &11754 & 7.99 & 21.58 & 70.43\\

\end{tabular}
\end{table}

\subsubsection{Influence of the secondary's inclination $i_b$ and the initial condition of the disk objects}
The result of all recorded collisions of the N-body simulations of the different binary star configurations are summarized in Table~\ref{tab:bin_confCollSum}, where we show the collision statistics for planar (all i00 configurations of Table~\ref{tab:binConfTP}) and i20-EIC (all i20-EIC configurations of Table~\ref{tab:binConfTP}) configurations. Table~\ref{tab:bin_confColl} in the appendix shows the breakdown of collisions for each configuration group individually. In both tables we distinguish between (i) accretion processes (perfect merging, graze and merge, partial accretion), (ii) destructive collisions (partial erosion, super catastrophic), and (iii) hit-and-run outcomes. The total number of collisions is quite similar for all groups of binary configurations: 18130 collisions in the planar configurations and 18017 collisions in the i20 configurations. The collisions are again divided into the three types of collision pairs, as mentioned above. Depending on the binary configuration, there are different proportions of collision outcomes for the respective collision pairs.\\
While in the planar cases all e-e collisions are accretive, in the i20-EIC configurations only $f_{\mathrm{accr}}\approx62\%$ are accretive. Destructive collisions are relatively low in all binary configurations: in the planar case $f_{\mathrm{destr}}\approx0\%$, in i20 configurations $f_{\mathrm{destr}}<1\%$. Hit-and-run collisions only play a major role in the inclined configurations with $f_{\mathrm{hr}}\approx37\%$.\\
The distribution changes slightly for e-p collisions. Accretive collisions are predominate again in the planar configurations, in the i20-EIC configurations $f_{\mathrm{accr}}\approx54\%$. The number of destructive collisions is similar to that of the e-e collisions: planar $f_{\mathrm{destr}}=0\%$, i20 configurations $f_{\mathrm{destr}}\approx0.67\%$. Hit-and-run collisions are also hardly present for the e-p collisions in the planar configurations ($f_{\mathrm{hr}}<1\%$). In the i20 configurations, however, they are almost half of the collisions ($f_{\mathrm{hr}}\approx45\%$).\\
For the p-p collisions, the distribution is different. In all binary configurations, hit-and-run collisions dominate (planar: $f_{\mathrm{hr}}\approx63\%$, i20 configurations: $f_{\mathrm{hr}}\approx70\%$). The deflections from the original disk plane and the additional velocity component are particularly responsible for the destructive collisions. In the i20-EIC configurations, the proportion of destructive collisions is already $f_{\mathrm{destr}}\approx21\%$. In the planar configurations, however, it remains low ($f_{\mathrm{destr}}<1\%$). Consequently, the ratio of accretive collisions also changes with $f_{\mathrm{accr}}\approx36\%$ in the planar configurations, and is significantly lower in the i20 configurations with $f_{\mathrm{accr}}\approx8\%$.\\
From Table~\ref{tab:bin_confCollSum} (taken the total number of p-p collisions $N_{\mathrm{coll}}$ into account) and Table~\ref{tab:bin_confColl} it can be seen that the majority of hit-and-run collisions come from p-p collisions, especially in the inclined configurations. In the case of destructive collisions almost all occur among planetesimals. On the other hand, a large part of the accretive collisions are e-p collisions. The reason for this is that the escape velocity depends on the total mass of the colliding objects \footnote{$V_{esc} = \sqrt{2GM_{tot}/(R_p+R_t)}$, with $M_{tot}$ the total mass of the colliding objects, and $R_p$ and $R_t$ the spherical radius of the impactor and target respectively} which is larger for more massive bodies. Consequently, the impact velocity, which is considered in units of the escape velocity, is lower and thus favours accretive collisions (taking into account the impact angle, of course). In addition, a smaller mass ratio between impactor and target ($\gamma = M_p/M_t$, $M_p$ mass of the impactor and $M_t$ mass of the target) leads to a higher possibility of accretive collision \citep[see Figure 11 of][]{LeiSte12}.\\
The distribution of collision outcomes differs between the EIC and PIC configurations. In the PIC configurations, there are far more p-p collisions (1429-1622), with hit-and-run collisions accounting for the largest share (820-1066). However, the tight configuration (a30-e02-i20-PIC) has more destructive collisions (233) and fewer accretive collisions (323) than the wider configurations (a100-PIC), with few destructive collisions (16-19) and a larger amount of accretive collisions (562-619). For the e-p collisions, less occurred in the PIC configurations (336-549) compared to the EIC configurations. The largest fraction accounts to the accretive collisions, with the tight configuration (a30-e02-i20-PIC) has the lowest amount (214), while the wider configurations (a100-PIC) have a larger faction of accretive collisions (501-549). Barely destructive collisions occur. The tight configuration shows a significant larger fraction of hit-and-run collisions (123) contrary to the wide configurations (26-34). The reason for the large fraction of accretive collisions of the wide PIC configurations (a100-PIC) compared to the wide EIC configurations, might be the effects mentioned in section \ref{subsec:PIC}. Due to the larger secular periods (Heppenheimer model) in the wider configurations, the disk objects exhibit lower eccentricities and inclinations over a longer time than in the tight configuration, resulting in lower impact velocities and thus more accretive collisions. The numbers of e-e collisions differ barely between the EIC and PIC configurations.

\section{Summary and Discussion}\label{discussion}
We investigated the dynamical evolution of planetary embryos -- planetesimal disks in various binary star configurations. The disk objects move in S-type orbits around the primary star, and the companion star (secondary), which is more or less distant and may be aligned or misaligned with the disk plane, disrupts the evolution of the disk -- depending on the parameters of the binary star. In particular, we investigated the influence of the planetesimal disk on the planetary embryos using N-body simulations. These calculations were very extensive, as the interactions of more than 2000 objects had to be calculated. With our self-developed GPU N-body code GANBISS, these calculations could be performed within a reasonable time. It is noteworthy that simulations that take into account the gravitational interactions of several thousand objects in binary star systems have not yet been carried out. Our research encompassed different binary configurations with variations in semi-major axis ($a_b$ from $30$ to $100\ \mathrm{au}$), eccentricity ($e_b$ from $0.0$ to $0.4$), and inclinations ($i_b=0{\degr}$ and $20{\degr}$). For the growth of terrestrial planets via collisions we used the so-called perfect merging approach where terrestrial planets or proto-planets could be formed in all systems within the computation time of $10\ \mathrm{Myrs}$. We therefore investigated the evolution of the disks only for the first $10$ million years of terrestrial planet formation, during which the interactions of a large number of objects must be calculated. Our simulations showed that most planetesimals merged with the planetary embryos, significantly reducing the number of objects in the various binary star configurations and meaning that the mass of the remaining planetesimals no longer had a significant influence on the evolution of the grown planets. After this period, the dynamic investigations of the systems can be continued over longer times (up to $100\ \mathrm{Myrs}$) using CPU N-body codes. Additionally, we compared different initial conditions for the planetary embryo-planetesimal disk in the misaligned binary star configurations, where we used either a dynamically cold disk (in planar and PIC configurations) or an already excited one (EIC configurations). In this context, we have seen that the misalignment of the secondary star with respect to the primary-disk plane is the critical factor for the final structure of a planetary system.\\
Our investigation revealed a slight inward migration of planetary embryos within the inclined configurations. The observed migration is based on the combined dissipative effects of dynamical friction and collisions with other disk objects. It should be noted that inward migration of planetesimals has also been detected in gas disks, but in that case of course to gas-induced friction \citep[see][]{Xieetal11}. In the post-gas phase, the migration acts as long as there are disk objects that can interact. As soon as they are lost due to collisions or ejections migration will stop. The initial inner boundary of the disk at $1\ \mathrm{au}$, coupled with the limited extent of inward migration, prevents collisions with the primary star, as observed in studies by \citet{Quietal02}. Consequently, no mass transfer from the disk towards the primary star is attributable to the inward migration in our studies. However, in close systems ($a_b=30\ \mathrm{au}$) we find some ejections of planetesimals and planetary embryos, due to the stronger influence of the secondary star. On the other hand, in the wide configurations (e.g.~a100-PIC) with an initially planar planetary embryo-planetesimal disk, the inward migration does not occur, due the longer periods on which the secular perturbation frequency act, and therefore the gravitational focusing is effective over a longer time. Our study confirms also an orbital alignment effect shown by \citet{Zhaetal18}. Their simulations have shown that the disk objects exhibit in $i_d$ up to $2i_b$. As long as planetesimals are present in the disk, they can dampen the maximum amplitudes of these oscillations.\\
A quick stability check of the planetary systems after $10\ \mathrm{Myrs}$ showed different results for the Hill and NAMD stability. While the Hill criterion yields stable, semi-stable and unstable systems in both planar and inclined configurations, the NAMD criterion estimates smaller values for all planar binary configurations and larger values for the inclined binary star configurations. Of course, the evolution of the computed systems is not after $10\ \mathrm{Myrs}$, and would take up to $100\ \mathrm{Myr}$ (which may also depend on the separation of the binary stars). As already mentioned before, due to the low number of remaining bodies, a CPU-based N-body integrator is sufficient to carry out the long term simulations. The final architecture of the systems will be investigated in a future study. Furthermore, the stability will be investigated in more detail considering the work of \citet{RodLai21} and longer integration times. $10\ \mathrm{Myrs}$ are generally too short for stability analyses.\\
Moreover, our post-processing analysis of the collisions has shown clearly that a more realistic consideration of the collisions might lead to different system configurations, especially in the inclined configurations, where the perturbation of the secondary and thus the effects on the impact parameters are noticeable. This is due to the large eccentricities and inclinations, which may increase the impact velocity. Larger impact velocities can have a negative effect on the formation of terrestrial planets. However, less destructive collisions occur in the PIC configurations, as most collisions happen at the beginning of the simulations, when eccentricities and inclinations of the disk objects are small, which increases the gravitational focusing and reduces the impact velocities. Besides the destructive collisions, a large amount of collision outcomes are hit-and-run collision. \\
A comparison of our post-processing analysis of collisions that occurred in the N-body simulations with previous works is not possible, as so far only similar studies have been published in single stars, see e.g.~those by \citet{SteLei12} or \citet{Quietal16}. In their studies, perturbations of giant planets (Jupiter and Saturn analogues) on the terrestrial planet formation were investigated while we examined planetary embryo-planetesimal disks in binary star systems. However, \citet{ChiMar22b} applied the study of \citet{LeiSte12} on N-body simulations in P-type motion, but did not examine the different collision outcomes in detail. \\
The evolution and so far formed systems in misaligned binary stars depend heavily on the initial conditions of the planet forming disk, especially in the wide binary configurations, which is due to the longer time-scale of the secular perturbations of the secondary star. However, the evolution of the disk in the intermediate (gas) phase in binary star systems is poorly understood and needs to be investigated further. This is beyond the scope of this paper and will be studied in the future.

\section{Conclusion}\label{conclusion}
The performed GPU-parallelized N-body simulations of embryo-planetesimal disks, containing some thousand objects in planar and misaligned binary star configurations showed that a misalignment between the binary plane and the planetary embryo - planetesimal disk plane has a strong influence in the evolution of the disk. Both an inward migration induced by collisions and dynamical friction, and an alignment of the inclination of the disk objects onto the inclination of the secondary star has been found. In principle terrestrial planets could form in misaligned binary star systems, but a deeper investigation of the collision outcomes among the disk objects show a large fraction of destructive events (especially in case of planetesimal-planetesimal collisions), which is not favourable for terrestrial planet formation. Thus, for binary systems with separations $>30-40\ \mathrm{au}$ terrestrial planet formation could be more difficult. A possible solution would be, as these destructive collisions create small fragments, which could be re-accreted by other disk objects \citep{Xieetal10,PaaLei10} \citep[][denoted it as lucky seeds]{Marthe19}. Thus, the late stage of terrestrial planet formation in a misaligned binary star system needs further investigations using an improved collision handling, including fragmentation \citep{Creetal21}. However, combining SPH and N-body simulations is computationally challenging. \citet{Buretal20} did it for a solar system-like architecture containing about 200 objects. The bottleneck in the simulations of \citet{Buretal20} was the N-body interactions. But in our study each N-body simulation contained more than 2000 objects, where a separate SPH simulation for each two-body-collision would be very time-consuming, thus computational expensive. The approach suggested in this investigation provides a more practicable solution for handling 2-body collisions occurring in N-body simulations. Thus, our approach is an improvement of the perfect merging method, for terrestrial planet formation in binary stars. Another approach for more realistic collisions outcomes could be a machine learning approach to predict the outcome of a collision \citep{Winetal23}. However, more detailed investigations are needed on the terrestrial planet formation especially in misaligned binary star systems. Our comparison of different initial conditions for the planet forming disk (initially dynamically cold and dynamically excited disks) has shown, that there is a strong influence on the evolution of the disk and the architecture of the planetary system especially in the binary star configurations.

\section*{Acknowledgements}
The authors want to thank the anonymous referee for useful comments that improved the manuscript. The authors want to acknowledge the support by the Austrian FWF - project PAT3059124 and P33351-N. The computational results presented have been achieved using the Vienna Scientific Cluster (projects 71637, 71686, 70320).

\section*{Data availability}
The data underlying this article are available in PHAIDRA, at \url{https://phaidra.univie.ac.at/detail/o:2174958}.

\bibliographystyle{mnras}
\bibliography{literature}

\appendix
  \onecolumn

\section{Final configurations}\label{app:A}

\begin{figure}
  \centering
  \includegraphics[width=1.0\textwidth]{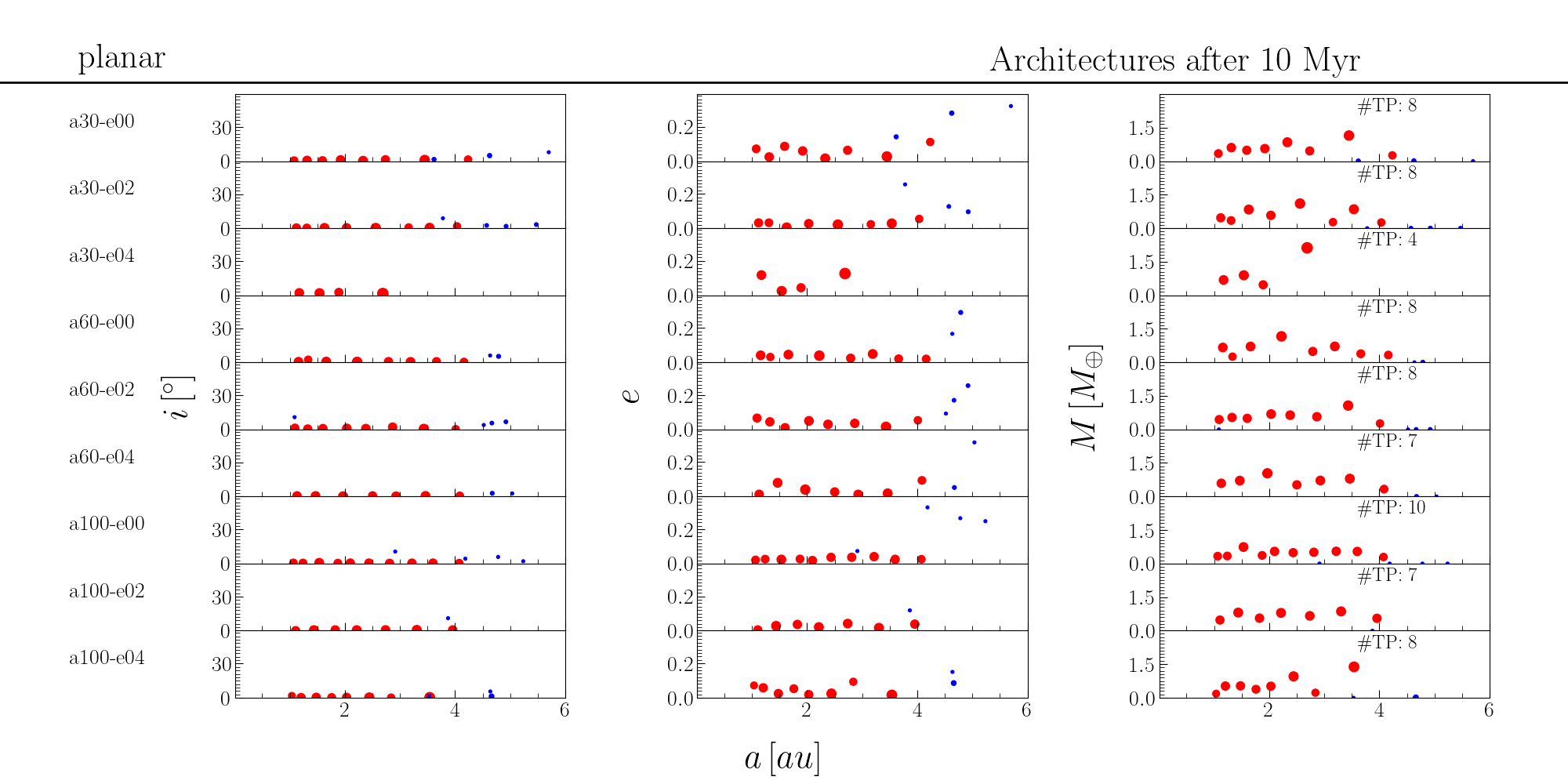}
  \caption{Architectures of all planar configurations after $10\ \mathrm{Myr}$ simulation time. The first column shows the initial conditions in $a$ and $i$ of the disk objects. Columns 2,3, and 4 show the final configuration in $a,i,e$ of the disk objects and their final mass for each computed planar configuration. Colours as in Fig.~\ref{fig:multiTStep30a02e20iE}. In the last column $\#TP$ shows the number of formed terrestrial planets for each configuration.}
  \label{fig:finConfPlan}
\end{figure}

\begin{figure}
  \centering
  \includegraphics[width=1.0\textwidth]{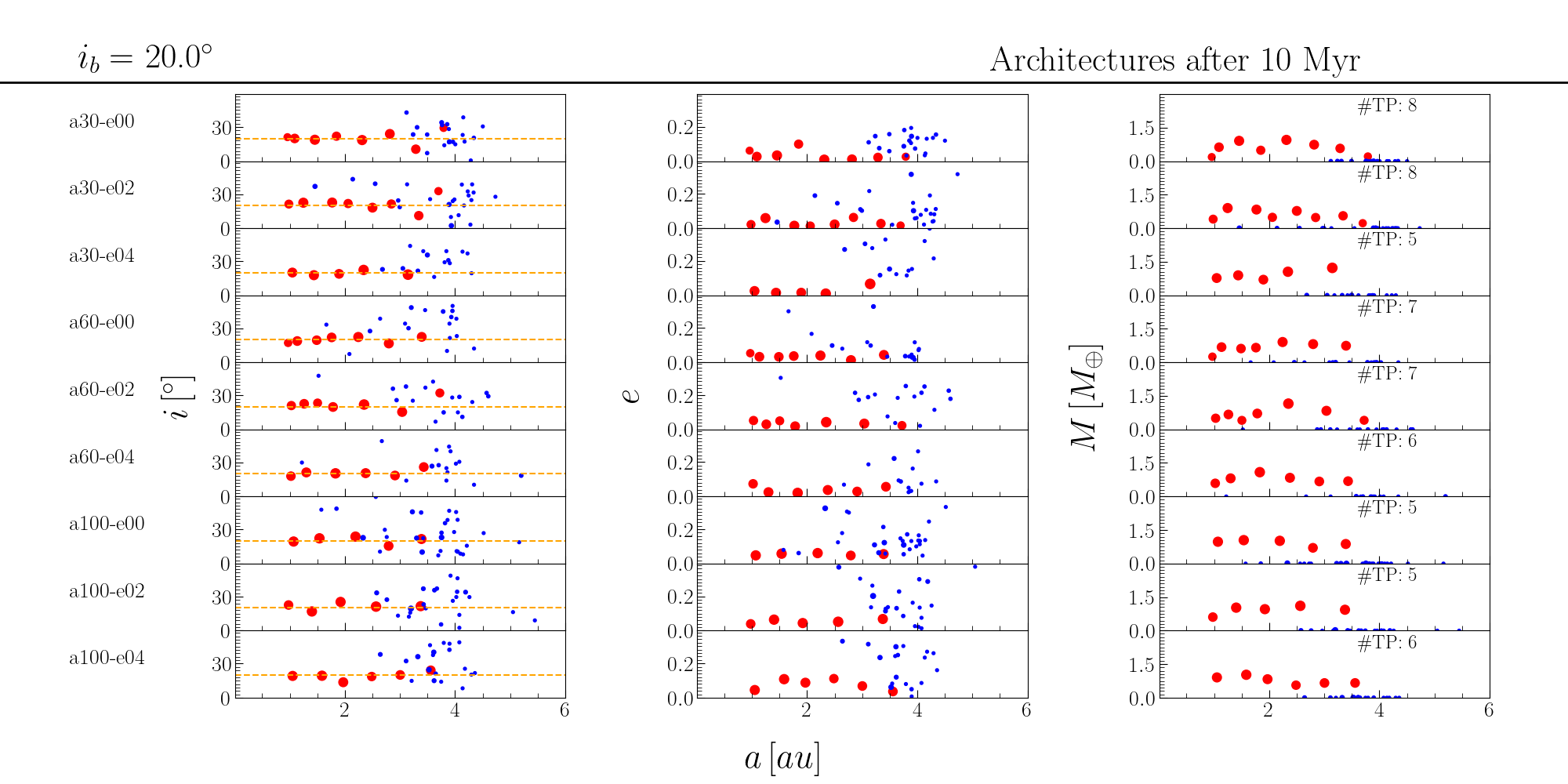}
  \caption{The same as in Fig.~\ref{fig:finConfPlan}, but for the i20-EIC configurations.}
  \label{fig:finConf20}
\end{figure}

\begin{figure}
  \centering
  \includegraphics[width=1.0\textwidth]{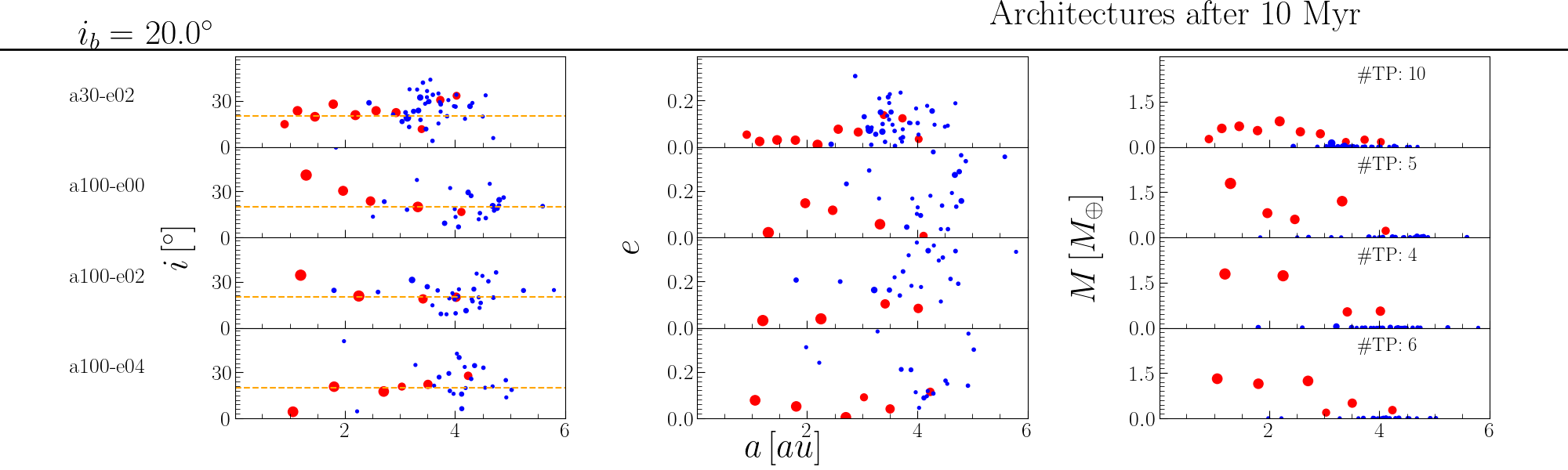}
  \caption{The same as in Fig.~\ref{fig:finConf20}, but for the PIC configurations.}
  \label{fig:finConfPIC}
\end{figure}
\clearpage
\pagebreak

  \section{Breakdown of collisions for each configuration}\label{app:B}

   \begin{table}
  \caption{The table shows the number of collisions for different collision partners and binary configurations. The first column lists each computed configuration. The second column shows the total number of collisions ($N_{\mathrm{coll}}$) in the given configuration. The next four columns show the number of collisions broken down into the respective outcomes for the various collision partners. Here $N_{\mathrm{coll}}$ is the total number of collisions, $N_{\mathrm{accr}}$ the accretive collisions (perfect merging, graze and merge, and partial accretion), $N_{\mathrm{destr}}$ the destructive collisions (partial erosion and super catastrophic), and $N_{\mathrm{hr}}$ the number of hit-and-run collisions.}
  \label{tab:bin_confColl}
  \centering
  \resizebox{\textwidth}{!}{\begin{tabular}{c | c | c c c c | c c c c | c c c c }
    & total & \multicolumn{4}{c|}{e-e} & \multicolumn{4}{c|}{e-p} & \multicolumn{4}{c}{p-p} \\
    \hline
    Configuration & $N_{\mathrm{coll}}$ & $N_{\mathrm{coll}}$ & $N_{\mathrm{accr}}$ & $N_{\mathrm{destr}}$ & $N_{\mathrm{hr}}$ & $N_{\mathrm{coll}}$ & $N_{\mathrm{accr}}$ & $N_{\mathrm{destr}}$ & $N_{\mathrm{hr}}$ & $N_{\mathrm{coll}}$ & $N_{\mathrm{accr}}$ & $N_{\mathrm{destr}}$ & $N_{\mathrm{hr}}$ \\
    \hline
    a30-e00-i00 & 2025 & 17 & 17 & 0 & 0 & 930 & 922 & 0 & 8 & 1078 & 391 & 1 & 686 \\
    a30-e02-i00 & 2012 & 17 & 17 & 0 & 0 & 885 & 879 & 0 & 6 & 1100 & 376 & 3 & 731 \\
    a30-e04-i00 & 2007 & 20 & 20 & 0 & 0 & 816 & 806 & 0 & 10 & 1171 & 398 & 4 & 769 \\
    a60-e00-i00 & 2015 & 17 & 17 & 0 & 0 & 930 & 923 & 0 & 7 & 1068 & 403 & 2 & 663 \\
    a60-e02-i00 & 2013 & 17 & 17 & 0 & 0 & 921 & 909 & 0 & 12 & 1079 & 393 & 3 & 679 \\
    a60-e04-i00 & 2016 & 18 & 18 & 0 & 0 & 916 & 911 & 0 & 5 & 1082 & 395 & 1 & 686 \\
    a100-e00-i00 & 2011 & 15 & 15 & 0 & 0 & 984 & 977 & 0 & 7 & 1012 & 396 & 3 & 613 \\
    a100-e02-i00 & 2017 & 18 & 18 & 0 & 0 & 992 & 984 &  0 & 8 & 1007 & 345 & 0 & 662 \\
    a100-e04-i00 & 2014 & 17 & 17 & 0 & 0 & 962 & 957 & 0 & 5 & 1035 & 371 & 1 & 663 \\
    \hline
    a30-e00-i20-EIC & 1997 & 18 & 11 & 0 & 7 & 692 & 354 & 2 & 336 & 1287 & 110 & 267 & 910 \\
    \textbf{a30-e02-i20-EIC} & 2004 & 18 & 13 & 0 & 5 & 686 & 360 & 3 & 323 & 1300 & 107 & 258 & 935 \\
    a30-e04-i20-EIC & 1995 & 20 & 12 & 0 & 8 & 656 & 360 & 6 & 290 & 1319 & 92 & 294 & 933 \\
    a60-e00-i20-EIC & 2006 & 18 & 13 & 0 & 5 & 666 & 345 & 6 & 315 & 1322 & 84 & 304 & 934 \\
    a60-e02-i20-EIC & 2006 & 18 & 13 & 0 & 5 & 645 & 322 & 5 & 318 & 1343 & 103 & 308 & 932 \\
    a60-e04-i20-EIC & 2009 & 21 & 14 & 0 & 7 & 636 & 332 & 4 & 300 & 1352 & 97 & 308 & 947\\
    \textbf{a100-e00-i20-EIC} & 1991 & 20 & 10 & 0 & 10 & 696 & 410 & 5 & 281 & 1257 & 117 & 261 & 897 \\
    \textbf{a100-e02-i20-EIC} & 2004 & 21 & 12 & 1 & 8 & 734 & 425 & 6 & 303 & 1249 & 114 & 270 & 865 \\
    \textbf{a100-e04-i20-EIC} & 2005 & 19 & 10 & 0 & 9 & 679 & 386 & 4 & 289 & 1307 & 115 & 267 & 925 \\
    \hline
    a30-e02-i20-PIC & 1976 & 18 & 11 & 0 & 7 & 336 & 214 & 2 & 123 & 1622 & 323 & 233 & 1066 \\
    a100-e00-i20-PIC & 1996 & 22 & 18 & 0 & 4 & 501 & 470 & 2 & 29 & 1473 & 619 & 17 & 837 \\
    a100-e02-i20-PIC & 1992 & 23 & 15 & 0 & 8 & 534 & 508 & 0 & 26 & 1435 & 562 & 19 & 854 \\
    a100-e04-i20-PIC & 1999 & 21 & 16 & 0 & 5 & 549 & 515 & 0 & 34 & 1429& 593 & 16 & 820 \\
  \end{tabular}}
  \end{table}

  \twocolumn

\bsp
\label{lastpage}

\end{document}